\documentclass[pra,twocolumn,superscriptaddress]{revtex4-1}
\usepackage[T1]{fontenc}
\usepackage[english]{babel}
\usepackage{siunitx}
\usepackage[utf8]{inputenc}
\usepackage{url}
\usepackage{dsfont}
\usepackage{bbm}
\usepackage{nicefrac}
\usepackage{setspace}
\usepackage[short]{optidef}
\usepackage{float}
\usepackage{comment}
\usepackage{dsfont}
\usepackage{amssymb}  
\usepackage[colorlinks=true, citecolor=blue,urlcolor=blue,linkcolor=blue,filecolor=black]{hyperref}
\usepackage[colorinlistoftodos, color=green!40, prependcaption]{todonotes}
\usepackage{ulem}
\usepackage{amsthm}
\usepackage{mathtools}
\usepackage{physics}
\usepackage{graphicx}
\usepackage{adjustbox}
\usepackage{placeins}
\usepackage{lipsum}
\usepackage{csquotes}

\newcommand{\beq}{\begin{equation}}
\newcommand{\eeq}{\end{equation}}
\renewcommand{\emph}{\textit}
\renewcommand{\tr}{\text{Tr}}
\usepackage{soul}
\usepackage{graphicx}% Include figure files
\usepackage{dcolumn}% Align table columns on decimal point
\usepackage{bm}% bold math

\usepackage[utf8]{inputenc}
\usepackage[T1]{fontenc}
\usepackage{mathptmx}

\begin{document}

\preprint{AIP/123-QED}

\title[]{Practical Semi-Device Independent Randomness Generation Based on Quantum State's Indistinguishability}
% Force line breaks with \\

\author{Hamid Tebyanian}
\thanks{These authors contributed equally to this work.}
\affiliation{Dipartimento di Ingegneria dell'Informazione, Universit\`a degli Studi di Padova, via Gradenigo 6B, IT-35131 Padova, Italy}

\author{Mujtaba Zahidy}
\thanks{These authors contributed equally to this work.}
\affiliation{Dipartimento di Ingegneria dell'Informazione, Universit\`a degli Studi di Padova, via Gradenigo 6B, IT-35131 Padova, Italy}

\author{Marco Avesani}
\affiliation{Dipartimento di Ingegneria dell'Informazione, Universit\`a degli Studi di Padova, via Gradenigo 6B, IT-35131 Padova, Italy}

\author{Andrea Stanco}
\affiliation{Dipartimento di Ingegneria dell'Informazione, Universit\`a degli Studi di Padova, via Gradenigo 6B, IT-35131 Padova, Italy}

\author{Paolo Villoresi}
\affiliation{Dipartimento di Ingegneria dell'Informazione, Universit\`a degli Studi di Padova, via Gradenigo 6B, IT-35131 Padova, Italy}
\affiliation{Padua Quantum Technologies Research Center, Universit\`a degli Studi di Padova, via Gradenigo 6B, IT-35131 Padova, Italy}
\affiliation{Istituto di Fotonica e Nanotecnologie, CNR, via Trasea 7, IT-35131 Padova, Italy}

\author{Giuseppe Vallone}
\affiliation{Dipartimento di Ingegneria dell'Informazione, Universit\`a degli Studi di Padova, via Gradenigo 6B, IT-35131 Padova, Italy}
\affiliation{Padua Quantum Technologies Research Center, Universit\`a degli Studi di Padova, via Gradenigo 6B, IT-35131 Padova, Italy}
\affiliation{Dipartimento di Fisica e Astronomia, Universit\`a degli Studi di Padova, via Marzolo 8, IT-35131 Padova, Italy}

\begin{abstract}
Semi-device independent (Semi-DI)
quantum random number generators (QRNG) gained attention 
{for security applications}, offering an excellent trade-off between security and generation rate. 
This paper presents a proof-of-principle time-bin encoding semi-DI QRNG experiments based on a prepare-and-measure scheme. 
The protocol requires two simple assumptions and a measurable condition: an upper-bound on the prepared pulses' energy. 
We lower-bound the conditional min-entropy from the energy-bound and the input-output correlation, determining the amount of genuine randomness that can be certified.
Moreover, we present a generalized optimization problem for bounding the min-entropy in the case of multiple input and outcomes, in the form of a semidefinite program (SDP).
The protocol is tested with a simple experimental setup, 
capable of realizing two configurations for the ternary time-bin encoding 
scheme. The experimental setup is easy-to-implement and comprises commercially available off-the-shelf (COTS) components at the telecom wavelength, granting a secure and certifiable entropy source. 
The combination of ease-of-implementation,
scalability, high security level and output-entropy, make our system a promising candidate for commercial QRNGs. 
\end{abstract}
\maketitle

\section{Introduction}
The world of cybersecurity is developing exceedingly fast, and the data encrypted by the traditional encryption methods are facing the danger of being revealed. 
Producing unpredictable and certified random numbers is a critical part of every cryptographic operation. There are many simple techniques to generate random numbers that rely on a deterministic phenomenon, however, these generators' security cannot be guaranteed since, in principle, they can always be predicted. On the contrary, quantum mechanics provides randomness based on its intrinsic behavior, which theoretically is an unpredictable source of secure random numbers \cite{Ma2016,Acin2016}. 

The most common approach to generate random numbers through a quantum process is by trusting the experiment's apparatus: these protocols are called trusted-device QRNGs. Trusted-device QRNGs are cheap, high-rate, and easy-to-implement \cite{Stanco2020,Regazzoni2021}, although the random numbers' security and privacy could be threatened \cite{QRNG_attack,QRNG_attack2}.
In fact, the behaviour of the trusted devices could deviate from the model and classical or quantum side-information could be leaked to the adversary's system, compromising the privacy of the numbers. Therefore, trust in the generator's devices can compromise the security of the system.
The highest level of security is offered by an approach called device-independent (DI) \cite{DI_Pironio_2010,DI_roger}. Considering there is no hypothesis on the devices' internal-working regularity, it is highly protected. However, this protocol's drawbacks are the low generation-rate and experimental complexity, making it less practical~\cite{Liu2021,Liu2018a,Zhang2020,DI_new,foletto2021experimental}. 

By introducing few assumptions on the working principles of the devices, it is possible to reduce the experimental complexity while increasing the generation rate; these protocols are called semi-DI \cite{Ma2016,Supic2020,tavakoli2021semideviceindependent}. The semi-DI scheme's assumptions can vary depending on users' needs, e.g.  source-DI~\cite{Ma_new1,avesani2020unbounded,Avesani2018, PhysRevX_SDI} have trusted measurement devices, or measurement-DI~\cite{MDI,MDI2}, where the
source device is 
trusted.
At the same time, there are protocols with weaker assumptions, e.g., bounding the state's overlap or  energy~\cite{tebyanian2020semidevice,Brask2017,Rusca2020,avesani2020}, granting a higher level of security.

In this work, by extending the approach proposed in~\cite{Brask2017} we demonstrate a semi-DI QRNG based on the ambiguity in discriminating non-orthogonal quantum states \cite{state_dis}.
Non-orthogonal quantum states can not be perfectly distinguished due to the inevitable uncertainty imposed by the quantum theory. This uncertainty can be exploited, as in this protocol, to generate secure and private random numbers.
A security estimation based on state overlap and unambiguous state discrimination was first derived in \cite{Brask2017, VanHimbeeck2017semidevice} and later implemented for coherent detection schemes in \cite{avesani2020,Rusca2020,tebyanian2020semidevice}.

We generalized the security framework initially presented in \cite{Brask2017}
in the case of a larger number of inputs and outputs (for more details and comparison, see Appendix \ref{app:c}).
We implement the protocol with a photonic setup based on a time-bin encoding with two configurations. In both configurations, we consider three inputs, while four and seven outcomes are tested in the respective structures.

The experimental setup is based on a prepare-and-measure scheme that features all-in-fiber commercially off-the-shelf (COTS) components at the telecom wavelength (1550 nm). The output entropy is evaluated given the correlation of the input-output data $p(b|x)$ along with the bound on the input states' energy that is the single measurable condition of this semi-DI QRNG. Furthermore, the user is capable of {monitoring} on-the-fly that the bound on the energy used to calculate the randomness rate is indeed verified by the given devices.
Note that we assume that the inputs are identically and independently distributed (I.I.D. hypothesis).

The reduced number of assumptions with respect to other types of semi-DI QRNG allows to reduce the trust in the employed devices, thus increasing its security, while keeping its performance on par with the commercial QRNGs \cite{gras2020quantum}. 
Finally, this implementation can be further miniaturized by integrating it directly on a chip as shown in \cite{QRNG_chip}. 

\section{ Framework }
\subsection{Protocol}
\begin{figure}[ht!]
\centering
\includegraphics[width=\linewidth]{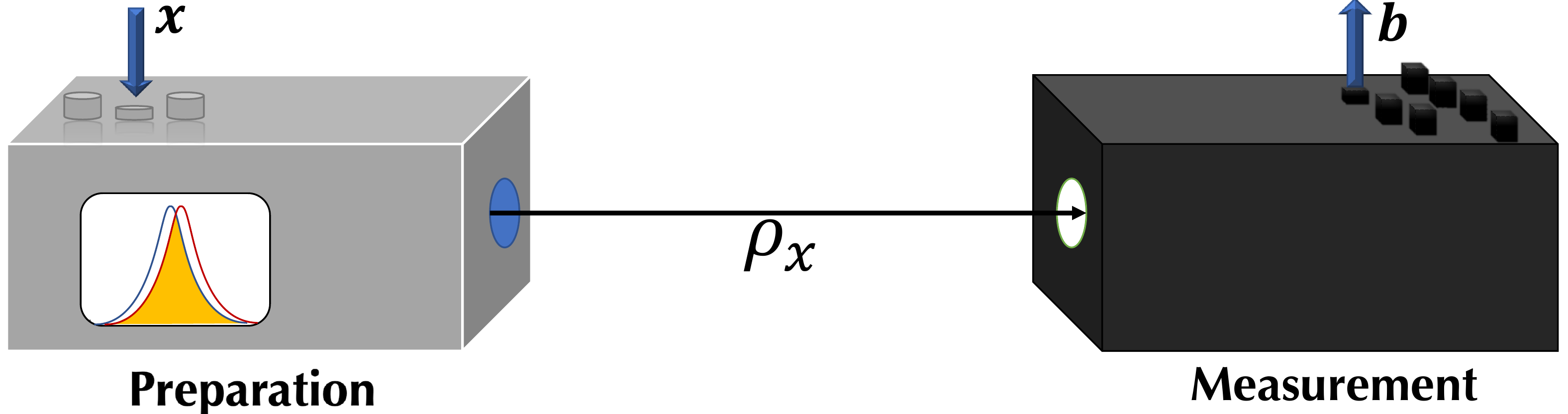}
\caption{ General schematic of the protocol; depending on the input $x$, the preparation device emits a quantum state $\rho_x$, with a single condition on the states' energy. The measurement site outputs $b$ after detecting the received states, as there is no assumption on the receiver side, it can be regarded as a black box.}
\label{fig:general}
\end{figure}

The experimental setup is based on a prepare-and-measure scheme, see Fig. \ref{fig:general}. A ternary input $x\in \{0,1,2\}$ is fed into the preparation device,
which prepares, accordingly,  a quantum state $\hat\rho_x$, that is sent to the measurement station. Here, after the measurement of the quantum state, the station returns an output $b\in 0,1,\cdots,d-1$.
The preparation and measurement devices are considered black boxes, with two simple assumptions 
on the preparation device: 
the prepared states are identically and independently distributed (I.I.D. hypothesis) and
no correlations between the preparation device and any external device are present. Randomness can be certified if the following  bound, 
easy-to-verify experimentally, holds on the energy of the prepared states:
\beq
\langle \hat N\rangle_{\rho_x}\leq \mu\,, \quad \forall x,
\label{energy_bound1}
\eeq
where $\hat N$ is the photon number operator 
(i.e. the energy of the state) and $\mu$ is its upper-bound.

If $\mu$ is below a certain threshold, the emitted states must be close to the vacuum, and so they must share some unavoidable overlap.
According to quantum mechanics,
non-orthogonal quantum states can not be deterministically distinguished
meaning that outcomes of any measurement cannot be predicted with certainty. 

From this simple idea it is possible to show that the amount of extractable randomness can be evaluated only by knowing the energy bound and the input-output correlations $P(b|x)$,
in a semi-DI way.
Indeed, the observation of certain correlations certifies that no pre-established strategies can fully reproduce the measured outcomes.
The values of the correlations allow to certify their quantum nature and allows to bound the amount of entropy in the outcomes. 

The scheme can be described as follows: the preparation device produces quantum states $\hat\rho_x$ while the measurement device performs a 
positive-operator valued measurement (POVM) $\hat \Pi_b^\lambda$. The classical variable $\lambda$, known to the adversary (e.g. the producer of the devices), represents the correlations between the measurement devices
and the adversary. Each different realization $\hat \Pi_b^\lambda$ labeled by $\lambda$ can be implemented with probability $p_\lambda$.
The input-output correlations $P(b|x)$ can then be written as
\beq
p(b|x) = \sum\limits_\lambda  {p_\lambda} 
\tr[\hat\rho_x \hat\Pi_b^\lambda],
\label{eq:constraints}
\eeq
In order to bound the amount of private randomness that can be certified we need to bound, the guessing probability $P_{\rm guess}$: the latter represents the maximum probability of guessing the outcome of the measurement device $b$ from the adversary point of view which
has full knowledge of the fundamental working principle of the experiment apparatus and the input $x$. $P_{\rm guess}$ can be evaluated as follows:
\beq
P_{\rm guess} =  \max \limits_{\{ p_\lambda ,\hat\rho_x,\hat\Pi _b^\lambda\}}
\left(\sum_{x} p_x \sum_{\lambda} p_\lambda 
\max_b \bigg\{ \Tr[\hat\rho_x \hat\Pi^\lambda_b] \bigg\}\right),
\label{eq:P_g}
\eeq
where $p_x$ is the probability of transmitting $x$.
We assume that the probability of sending different inputs ($x$) is balanced $p_x=\frac13$.
The overall maximization is performed on the states and operators $\{ p_\lambda ,\hat\rho_x,\hat\Pi _b^\lambda\}$ that are compatible with the observed correlations and thus satisfy the constraint of Eq. ~\eqref{eq:constraints}.

Following the same approach shown in \cite{Brask2017},
since
the preparation device shares no correlation with the environment, the maximum  $P_{\rm guess}$ is achieved when the states $\hat\rho_x$ are pure states, $\hat\rho_x=\ket{\psi_x}\bra{\psi_x}$.
Since the energy bound~\eqref{energy_bound1} implies on pure states a bound on their overlap (see~\cite{Himbeeck2019CorrelationsConstraints,avesani2020}) $|\braket{\psi_x}{\psi_y}|\geq1-2\mu\equiv\delta$, the choice that maximize  $P_{\rm guess}$ is obtained when the bound is saturated, namely $|\braket{\psi_x}{\psi_y}|=\delta$, $\forall x,y$.
Then, without losing generalities, the three states $\ket{\psi_x}$ can be then written as a linear combination of three orthonormal states $\ket{0}$,
$\ket{1}$, $\ket{2}$ as follows:
\begin{equation}
\begin{aligned}
    \ket{\psi_0}&=\ket0\,,
    \\
    \ket{\psi_1}&=\delta \ket0+
    \sqrt{1-\delta^2}\ket1\,,
    \\
    \ket{\psi_2}&=\delta \ket0+
    \delta \sqrt{\frac{1-\delta}{1+\delta}}\ket1 +
    \sqrt{\frac{1+\delta-2\delta^2}{1+\delta}}
    \ket2 
\end{aligned}
\label{state:ternary}
\end{equation}
while  $P_{\rm guess}$ can be written as
\beq
 P_{\rm guess} = \frac{1}{3} \max \limits_{\{ p_\lambda ,\Pi _b^\lambda\}} 
\left( \sum\limits_{x = 0}^{2} \sum\limits_{\lambda}^{d-1} {p_\lambda } \max\limits_{b}  \bigg[ \bra{\psi_x}\Pi^\lambda_b\ket{\psi_x} \bigg] \right)
\label{eq:P_g2}
\eeq
It is possible to cast Eq. (\ref{eq:P_g2}) into an semi-definite programming (SDP) problem, which can be efficiently solved (see appendix \ref{app:a}).
By inserting the input-output correlations $p(b|x)$ into the SDP, we can obtain a bound $P_{\rm g}$ on the guessing probability  and the conditional min-entropy\cite{entropy}
that quantifies the amount of private randomness 
\beq
\label{eq:Hmin}
{
H_{\rm min}=-\log_2 \{ P_{g}\}.
}
\eeq
\begin{figure}[t!]
\includegraphics[width=\linewidth]{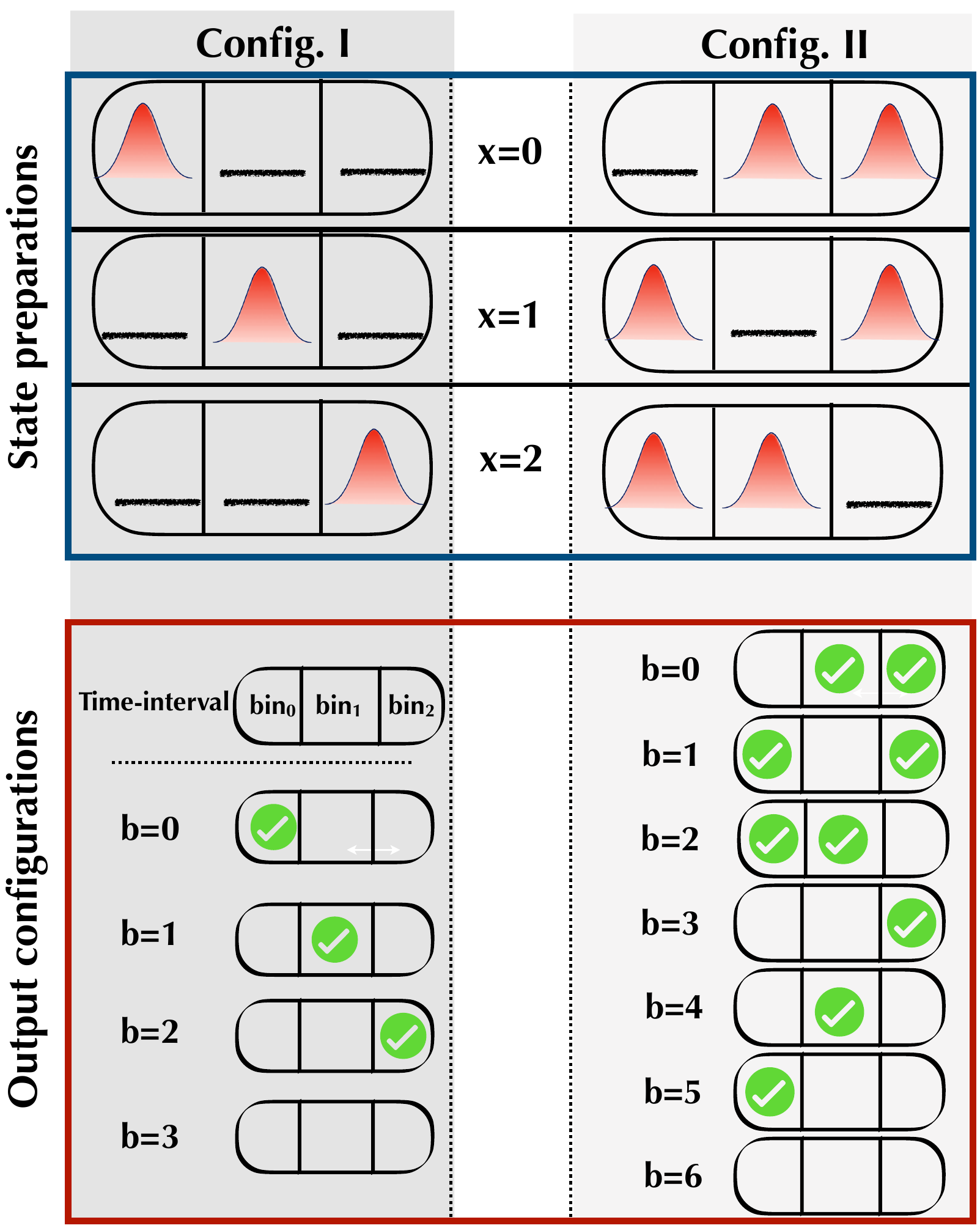}
\caption{ \textit{Blue-box:} Proposed states preparation configurations; in Config. I, there is one weak-coherent state in each time-bin, while the second configuration owns two weak-coherent states per time-bin. \textit{Red-box:} possible detection outcomes for the Config. I (left-side), and Config. II (right-side).}
\label{fig:conf1}
\end{figure}
Finally, after obtaining a bound on the min-entropy, secure and private random numbers can be obtained thanks to the Leftover Hashing Lemma, using a Toeplitz  randomness extractor~\cite{Tomamichel2011}.
\subsection{Implementation}
{The semi-DI protocol with ternary inputs and multiple outcomes described in the previous section can be implemented in different ways. In this work we present}
 two configurations based on the ternary time-bin encoding 
 {shown in the top box of Fig. \ref{fig:conf1}}. In the first configuration (Config. I), the transmitter emits a coherent state $\ket{\alpha}$ once every three bins,
 {while in the other two time-bins} the vacuum state is present. In contrast, in the second configuration (Config. II), the vacuum state and weak coherent pules are reversed. 
 For both configurations
we choose $\mu=|\alpha|^2$ such that the condition written in eq. \eqref{energy_bound1}
is satisfied.

The main advantage of such implementations is the low experimental complexity of the state's preparation and the possibility to easily monitor the energy of the prepared states.
{For the first configuration (Config. I), shown in Fig. \ref{fig:conf1} (lower box),} four possible outcomes $b \in \{0,1,2,3\}$ are considered, where $b=0$, $b=1$, and $b=2$ occur when a detection is registered in the early ($bin_0$), middle ($bin_1$), and late ($bin_2$) time-interval, respectively,  and if no click or more than one click
is recorded, then the outcome is $b=3$.
\begin{figure*}[htb!]
\centering
\includegraphics[width=6.3in]{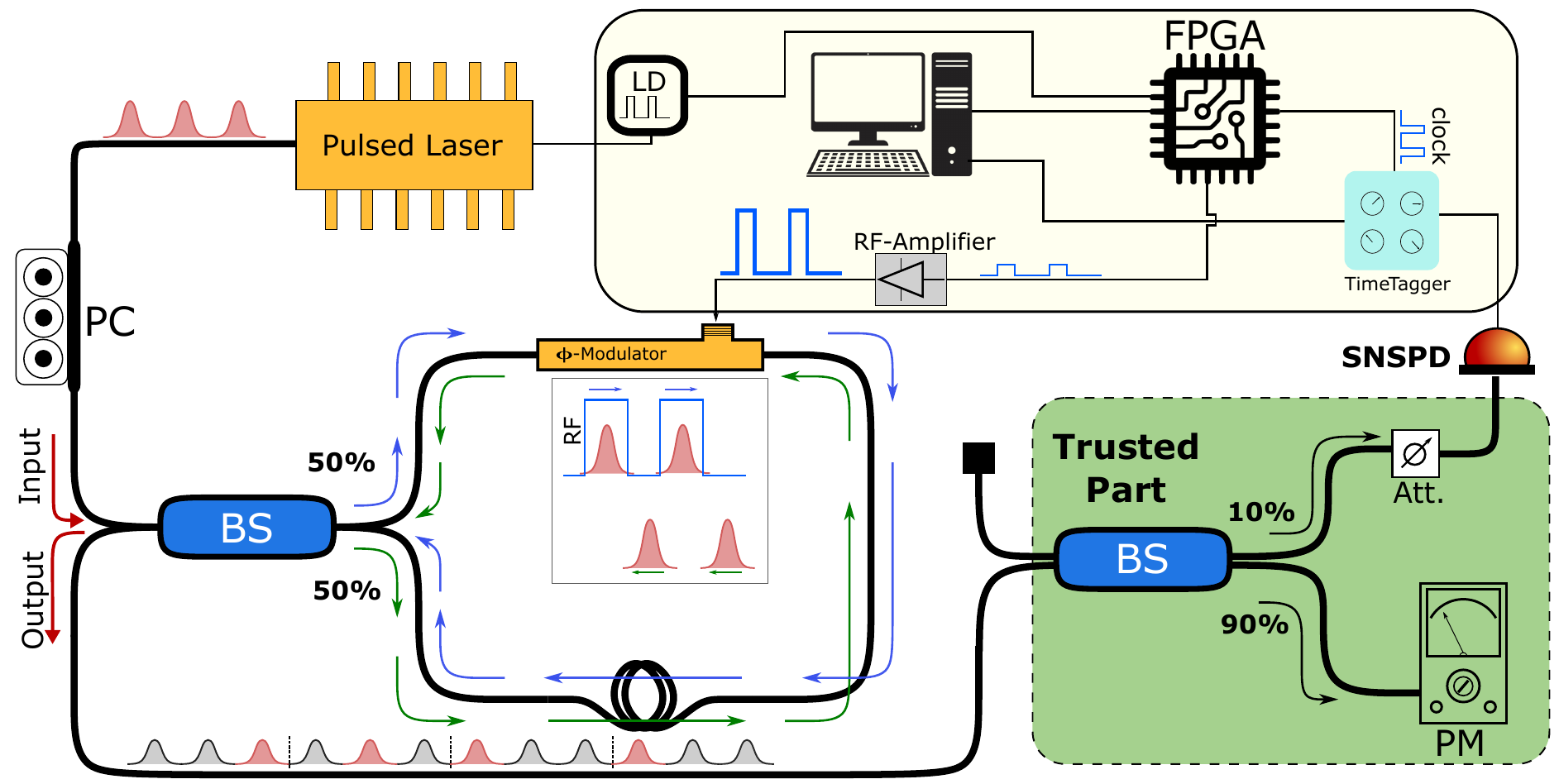}
\caption{Experimental setup: A pulsed laser emits pulses at 1550 nm to a polarization controller (PC) and then a
Sagnac interferometer (SI).
One path in
the SI is experiencing either an extra $0$ or $\pi$-phase shift with respect to the other one. The two parts interfere and recombine at the beam-splitter (BS). Depending on the phase shift, light is redirected to either output or back to the input. Later the single photons are detected with a single-photon detector, in this case, an SNSPD. A time to digital converter (TDC) converts the SNSPD detection event to time-stamps which are analyzed in post-processing. A field-programmable gate array (FPGA) provides the electrical signal to drive the laser driver (LD), phase modulator, and synchronization clock.}
\label{fig:setup}
\end{figure*}
On the other hand, for the second configuration (Config. II), 
a larger number of outcomes are possible. 
Let's for instance consider the case where $x=0$ is chosen for Config. II (see Fig. \ref{fig:general}).
Due to the low values of $\alpha$ imposed by the energy bound and the non-unity efficiency of the detectors, it is possible that only one of the two pulses is detected ($b=3$ or $b=4$ in Fig. \ref{fig:conf1}), or no pulses at all ($b=6$ in Fig. \ref{fig:conf1}). Thus, the total number of outcomes is increased from four to seven, with respect to the previous configuration.

\subsection{Input-output Correlation}
Depending on the input $x$, the transmitter sends one of the ternary states represented in Fig. \ref{fig:general}. We underline that the input $x$ are identically distributed and independent from the devices.
The states are measured at the receiver through a single-photon detector, in this case, a superconducting nanowire single-photon detector (SNSPD) \cite{SNSPD}. Based on the detection events and their
{arrival times}, the receiver outputs $b \in \{0,1,2,3\}$, or $\{0,1,2,3,4,5,6\}$.
{Given} the inputs $x$, and outputs $b$, we can compute the input-output correlation of the measurement and preparation devices $p(b|x)$, namely the probability of obtaining outcome $b$ given the input $x$.

In practice, the experimental setup is always combined with imperfections, mainly originated from the experimental apparatus; therefore, considering an
ideal measurement
would over-simplify our detection model. For example, the detector's dark count, background noise or imperfections in the state preparation, could lead to a theoretically impossible detection event. 
Therefore, we take these effects into account by introducing a value $\epsilon$ associated with the noise. We point out that the parameter $\epsilon$ is only useful for a correct modeling of the expected experimental probabilities, but it is not used in the $P_{\rm guess}$ evaluation and it has no impact on the security and performances of the protocol. 

The models used to describe the conditional probabilities $p(b|x)$ are the following:

\uline{Config. I}, $\forall x \in \{0,1,2\}$
\beq
\begin{aligned}
& p(b=x|x)= (1-\xi+\xi\epsilon)(1-\epsilon)^2\,, \\
& p(b \neq x\wedge b\neq 3|x)= \xi\epsilon (1-\epsilon)^2\,,
\\
& p(b=3|x)=  1-p(b\neq 3|x)\,, \\
\end{aligned}
\label{prob:congig 1}
\eeq
where $\xi = |\bra{\alpha}\ket{0}|^2=e^{-|\alpha|^2}$.

\uline{Config. II}, $\forall x \in \{0,1,2\}$
\beq
\begin{aligned}
& p(b=x|x)= (1-\xi+\xi\epsilon)^2(1-\epsilon) \,,
\\
& p(b=\varnothing_x|x)=  (1-\xi+\xi\epsilon)\xi(1-\epsilon)^2\,,
\\
& p(b=\varnothing''_x|x)=  \epsilon\xi^2(1-\epsilon)^2 \,,
\\
& p(b \neq x\wedge b<3|x)= (1-\xi+\xi\epsilon) \epsilon\xi(1-\epsilon) \,,
\\
& p(b=6|x)=  1-p(b\neq 6|x) \,.
\end{aligned}
\label{prob:congig 2}
\eeq
where $\varnothing' \in \{b=6\}$, $\varnothing_0 \in \{b=3,b=4\}$, $\varnothing_1 \in \{b=3,b=5\}$, $\varnothing_2 \in \{b=4,b=5\}$, $\varnothing''_0 \in \{b=5\}$, $\varnothing''_1 \in \{b=4\}$, and $\varnothing''_2 \in \{b=3\}$. Inserting these probabilities to the SDP (Eq. \ref{eq:SDP_primal}), we can compute the expected achievable min-entropy $H_{min}$ with our system. 

The advantage of this scheme 
compared with other solutions based on coherent detection is the simplicity 
of the experimental setup, which
does not require any 
complex phase-correction 
stabilization or further post-processing.
These advantages are particularly relevant for real-time implementations.
On the other hand, the possible drawback could be the random number generation rate, which, compared with similar continuous-variable systems \cite{avesani2020,Rusca2020}, is drastically lower, due to the high dead-time of the current SPDs~\cite{SPD}. 

\section{Experiment}

The experimental setup is depicted in Fig. \ref{fig:setup}. The realization is based on an all-in-fiber scheme with components that are commercially available off-the-shelf (COTS). The setup's core is a fast and self-stabilized optical switch based on Sagnac interferometer (SI)\cite{Roberts:18}, capable of operating up to GHz range. 
The switch is comprised of a ($50:50$) polarization-maintaining (PM) fiber-beamsplitter (BS), PM fiber delay line and a LiNbO\textsubscript{3} phase modulator (MPZ-LN-20 by iXblue). The ($50:50$) BS is used to split a pulse in two that travel in the Sagnac loop clockwise (CW) and counter-clockwise (CCW). The phase modulator applies a $0$ or $\pi$-phase shift to the CW pulse while leaving the CCW one intact. The two parts are then recombined again at the BS and according to the phase modulation value are either redirected to the trusted part and then measurement unit or send back toward the laser where it is blocked by the internal isolator.
\begin{figure*}[htb!]
\centering
\includegraphics[width=\linewidth]{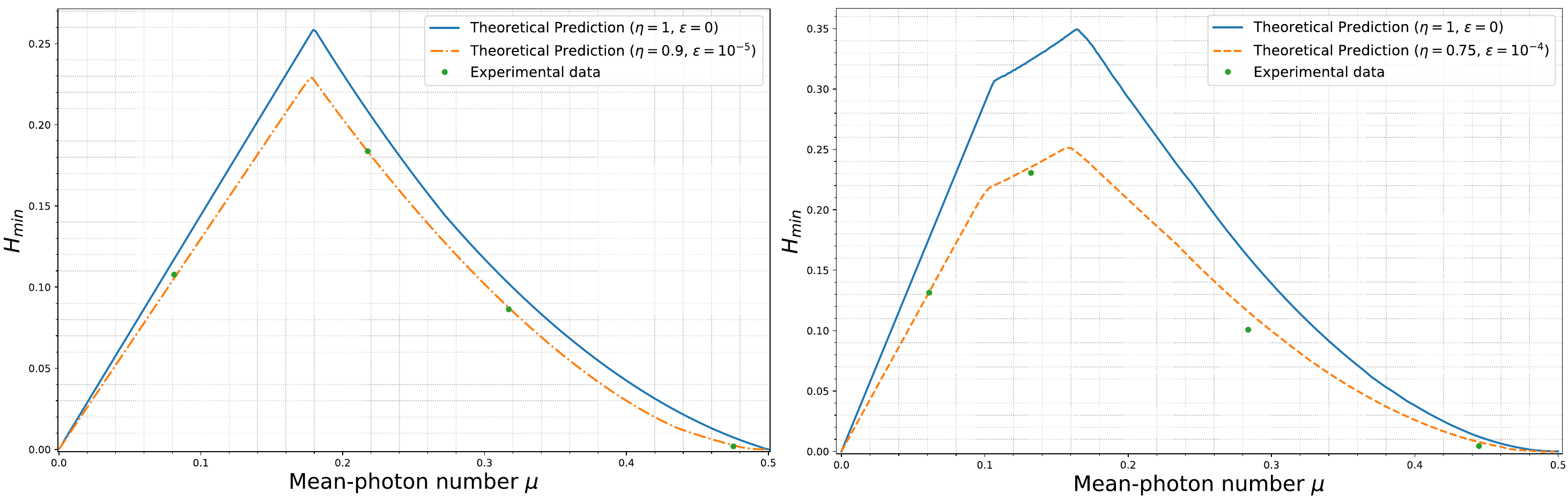}
\caption{The conditional min-entropy as a function of the mean-photon number $\mu$ for Config. I (left-side), and Config. II (right-side). The dashed and solid-line curves show the theoretical prediction with and without the experimental loss, respectively. The green dots represent the experimental data.
An SNSPD with detection efficiency equal to $75\%$ is used for Config. II, while an SNSPD with higher detection efficiency, $90\%$, is used for the Cofing. I.}
\label{fig:entropy}
\end{figure*}
The main advantage of the self-compensating 
Sagnac implementation over other types of intensity 
modulators is its resilience against phase fluctuations, ensuring very high extinction ratio at the output as well as high speed and long-term stability.
Unlike other intensity modulator this device does not requires to be stabilized in temperature or bias voltage.

A pulsed laser emitting at 1550 nm with ~2 ns pulse-width and fixed repetition rate of 10 MHz generates the train of pulses, which is first sent to a polarization controller (PC) and then to the input port of the switch. The input power is controlled accurately by changing its polarization via the polarization controller, where the PM-fiber BS acts as a polarizer. The output port of the switch is connected to a ($90:10$) PM-fiber BS, where the $90\%$ output is used to monitor the power and the $10\%$ is further transmitted along the optical path for the randomness generation.

A field programmable gate array (FPGA) board (ZedBoard by Avnet) provides the electrical signals to trigger the laser driver (LD) as well as phase modulation and a clock signal to synchronize the events. The phase modulation signal is amplified with an RF amplifier and then is used to drive the phase modulator.
States $\rho_x$ are generated by properly switching the input pulse, removing two (one) from every three pulses of the pulse train in Config. I (II). A typical output of the optical switch for Config. I is depicted in Fig. \ref{fig:setup}. An arbitrary sequence can be fed into the FPGA to perform the switching.
Two sequences of randomly distributed states, according to Config. I and II, are created and used for the experiment.    

Finally, the mean photon number at the exit of the transmitter is regulated with extra attenuation (Att.) set properly at the beginning of the experiment and is left fixed to maintain the ratio of the output power and monitor.
Prior to each run, the power is monitored and the average mean photon number per pulse is registered for the SDP and post-processing stages. 

For the measurement, we exploited SNSPDs with different detection efficiencies to inspect matching of the results with the theoretical predictions for each configuration. Further analysis of the performances as a function of the detection efficiency is contained in Appendix~\ref{app:c}. The very low dark count and dead-time of SNSPDs allow for measurement and symbol detection at high repetition rates where, for example, $\mu$-second range hold-off time of single-photon avalanche diodes (SPAD) limits the detection rate to tens of kilo-symbols per second.
Detection events are tagged with a time-to-digital converter (TDC) and the data is sent to a computer for post-processing. 
From the set of detections $b$ and the string of input $x$, it was possible to obtain the experimental conditional probabilities $p(b|x)$ for both configurations shown in Fig. \ref{fig:conf1}. 
\section{Results}
This section presents the 
results obtained from the experimental data.
We compare this experimental results with the model given by Eq. \eqref{prob:congig 1} and Eq. \eqref{prob:congig 2}
After implementing the experimental setup, represented in Fig. \ref{fig:setup}, we performed several measurement-runs with various mean-photon numbers $\mu$. The mean-photon number is determined by a calibrated optical powermeter per operation run, represented in the green box in Fig. \ref{fig:setup}. 
Collecting the receiver's outcomes $b$, and given the input sequence $x$, we calculate the input-output correlation $p(b|x)$. The extractable amount of randomness is then estimated by inserting $p(b|x)$ and $\mu$'s experimental values into the SDP code.

Fig. \ref{fig:entropy} shows the conditional min-entropy 
per measurement, as a function of the mean-photon number for the two supported configurations.
The experimentally obtained error value for Config. I and II are $\epsilon=10^{-5}$, and $\epsilon=10^{-4}$ respectively.
The difference in the 
$\epsilon$ values is due to the switch performance, noise and dark count rate (DCR) of the detectors which are $\simeq70$cps and $\simeq1400$cps in free-running, respectively. 

This shows an excellent stability and performance of the switch as well as the detectors.
In both plots, the blue curve represents the theoretical prediction without considering detection loss (perfect detector), while in the dashed orange curve, the losses (e.g., detector's efficiency) are also considered.
The green dots correspond to the experimental data obtained with two SNSPDs with different efficiencies; $90\%$ (used for Config. I), and $75\%$ (used for Config. II). Comparing the experimental data and theoretical predictions, we see an excellent agreement between them. 

From the theoretical model,
the maximum conditional min-entropy with a lossless detector is 0.258, and 0.349 bits per measurement {for Config. I and II, respectively}. They occur when the mean-photon number $\mu$ is roughly around $0.18$, and $0.164$ for Config. I and II.

Nevertheless, taking the losses into consideration, the conditional min-entropy recedes from its optimum value. Indeed the output entropy is very sensitive to the detector efficiency and losses. A comprehensive study of the amount of extractable randomness versus detectors' efficiency for two different assumptions (energy and overlap) is presented in Appendix \ref{app:c}.
Taking into account the parameters $\eta$ and $\epsilon$ that model our experiment, the maximum min-entropy that can be achieved experimentally are  0.183 and 0.23 for Config. I and II, respectively. We point our that the two configurations were not tested experimentally at their optimal points, but we tested they systems for some $  \mu$ values as a proof-of-principle demonstration.

Exploiting an optical switch rather than modulating the pulses directly on the laser has the advantage of avoiding fluctuations in mean photon number per pulse at the source due to laser cavity relaxation time. 
Besides, implementing binary or ternary states and states with higher number of time bins, e.g., $4$, $5$, etc., and various configurations can be readily done with this experimental setup, provided that the input to the FPGA is modified accordingly. Appendix \ref{app:c} compares the conditional min-entropy for several time-bins strategies.

Finally, it should be noted that this is a proof-of-principle experiment, and it can be significantly improved and optimized in forthcoming works, particularly by utilizing integrated photonics.

\section{conclusion}
In conclusion, we have presented a practical 
semi-DI QRNG based on
{ternary input} and measurements with multiple outcomes. 
Furthermore, we showed that it is possible to
{realize} two different implementations with a
{simple} setup based on
time-bin encoding. In addition, we compared our results with a binary modulated system and showed that by increasing the number of inputs from two to three, the output randomness increases accordingly. 
The proposed protocol features an increased security with respect to common QRNG, since it only requires two simple assumptions and a measurable condition on the prepared pulses' energy. 
The latter condition is experimentally easier to verify respect to other semi-DI protocol, for example based on an overlap bound.
Simultaneously, the protocol is practical,
since it can be implemented with a simple all-fiber optical setup at telecom wavelength with only commercial off-the-shelf components.
The performances of this proof-of-principle implementation could be further increased using faster repetition rates, faster modulation or integrated optics.

The proposed setup can also be useful to test higher 
dimensional states from an experimental point of view. 
In fact, this implementation only requires binary electrical signals even for higher dimensional states, while coherent systems require multi-amplitude modulations, increasing the complexity of the driving electronics.
Compared to the security estimation presented in \cite{Himbeeck2019CorrelationsConstraints}, our security evaluation requires an additional assumption (I.I.D hypothesis). Nevertheless, our protocol can be readily generalized for more input-outcome cases, 
while it is not clear how the security estimation provided in \cite{Himbeeck2019CorrelationsConstraints} can be generalized for more input and outputs.
Indeed, one of the main objectives of semi-DI protocols is to facilitate the implementation and improve the generation rate while keeping the security relatively high, which is contemplated in our protocol.
To conclude, our work shows how the increased number of input and output can improve the secure generation rate of QRNG in the semi-DI framework for future devices with simple experimental setups and high-security levels.

\begin{acknowledgments}
This work was supported by: ``Fondazione Cassa di Risparmio di Padova e Rovigo'' with the project QUASAR funded within the call ``Ricerca Scientifica di Eccellenza 2018''; 
MIUR (Italian Minister for Education) under the initiative ``Departments of Excellence'' (Law 232/2016); EU-H2020 program under the Marie Sklodowska Curie action, project QCALL (Grant No. GA 675662).
\end{acknowledgments}
\bibliography{bibliography}

\begin{thebibliography}{37}%
\makeatletter
\providecommand \@ifxundefined [1]{%
 \@ifx{#1\undefined}
}%
\providecommand \@ifnum [1]{%
 \ifnum #1\expandafter \@firstoftwo
 \else \expandafter \@secondoftwo
 \fi
}%
\providecommand \@ifx [1]{%
 \ifx #1\expandafter \@firstoftwo
 \else \expandafter \@secondoftwo
 \fi
}%
\providecommand \natexlab [1]{#1}%
\providecommand \enquote  [1]{``#1''}%
\providecommand \bibnamefont  [1]{#1}%
\providecommand \bibfnamefont [1]{#1}%
\providecommand \citenamefont [1]{#1}%
\providecommand \href@noop [0]{\@secondoftwo}%
\providecommand \href [0]{\begingroup \@sanitize@url \@href}%
\providecommand \@href[1]{\@@startlink{#1}\@@href}%
\providecommand \@@href[1]{\endgroup#1\@@endlink}%
\providecommand \@sanitize@url [0]{\catcode `\\12\catcode `\$12\catcode
  `\&12\catcode `\#12\catcode `\^12\catcode `\_12\catcode `\%12\relax}%
\providecommand \@@startlink[1]{}%
\providecommand \@@endlink[0]{}%
\providecommand \url  [0]{\begingroup\@sanitize@url \@url }%
\providecommand \@url [1]{\endgroup\@href {#1}{\urlprefix }}%
\providecommand \urlprefix  [0]{URL }%
\providecommand \Eprint [0]{\href }%
\providecommand \doibase [0]{http://dx.doi.org/}%
\providecommand \selectlanguage [0]{\@gobble}%
\providecommand \bibinfo  [0]{\@secondoftwo}%
\providecommand \bibfield  [0]{\@secondoftwo}%
\providecommand \translation [1]{[#1]}%
\providecommand \BibitemOpen [0]{}%
\providecommand \bibitemStop [0]{}%
\providecommand \bibitemNoStop [0]{.\EOS\space}%
\providecommand \EOS [0]{\spacefactor3000\relax}%
\providecommand \BibitemShut  [1]{\csname bibitem#1\endcsname}%
\let\auto@bib@innerbib\@empty
%</preamble>
\bibitem [{\citenamefont {Ma}\ \emph {et~al.}(2016)\citenamefont {Ma},
  \citenamefont {Yuan}, \citenamefont {Cao}, \citenamefont {Qi},\ and\
  \citenamefont {Zhang}}]{Ma2016}%
  \BibitemOpen
  \bibfield  {author} {\bibinfo {author} {\bibfnamefont {X.}~\bibnamefont
  {Ma}}, \bibinfo {author} {\bibfnamefont {X.}~\bibnamefont {Yuan}}, \bibinfo
  {author} {\bibfnamefont {Z.}~\bibnamefont {Cao}}, \bibinfo {author}
  {\bibfnamefont {B.}~\bibnamefont {Qi}}, \ and\ \bibinfo {author}
  {\bibfnamefont {Z.}~\bibnamefont {Zhang}},\ }\href {\doibase
  10.1038/npjqi.2016.21} {\bibfield  {journal} {\bibinfo  {journal} {npj
  Quantum Information}\ }\textbf {\bibinfo {volume} {2}},\ \bibinfo {pages}
  {16021} (\bibinfo {year} {2016})}\BibitemShut {NoStop}%
\bibitem [{\citenamefont {Ac{\'{i}}n}\ and\ \citenamefont
  {Masanes}(2016)}]{Acin2016}%
  \BibitemOpen
  \bibfield  {author} {\bibinfo {author} {\bibfnamefont {A.}~\bibnamefont
  {Ac{\'{i}}n}}\ and\ \bibinfo {author} {\bibfnamefont {L.}~\bibnamefont
  {Masanes}},\ }\href {\doibase 10.1038/nature20119} {\bibfield  {journal}
  {\bibinfo  {journal} {Nature}\ }\textbf {\bibinfo {volume} {540}},\ \bibinfo
  {pages} {213} (\bibinfo {year} {2016})}\BibitemShut {NoStop}%
\bibitem [{\citenamefont {Stanco}\ \emph {et~al.}(2020)\citenamefont {Stanco},
  \citenamefont {Marangon}, \citenamefont {Vallone}, \citenamefont {Burri},
  \citenamefont {Charbon},\ and\ \citenamefont {Villoresi}}]{Stanco2020}%
  \BibitemOpen
  \bibfield  {author} {\bibinfo {author} {\bibfnamefont {A.}~\bibnamefont
  {Stanco}}, \bibinfo {author} {\bibfnamefont {D.~G.}\ \bibnamefont
  {Marangon}}, \bibinfo {author} {\bibfnamefont {G.}~\bibnamefont {Vallone}},
  \bibinfo {author} {\bibfnamefont {S.}~\bibnamefont {Burri}}, \bibinfo
  {author} {\bibfnamefont {E.}~\bibnamefont {Charbon}}, \ and\ \bibinfo
  {author} {\bibfnamefont {P.}~\bibnamefont {Villoresi}},\ }\href {\doibase
  10.1103/PhysRevResearch.2.023287} {\bibfield  {journal} {\bibinfo  {journal}
  {Phys. Rev. Research}\ }\textbf {\bibinfo {volume} {2}},\ \bibinfo {pages}
  {023287} (\bibinfo {year} {2020})}\BibitemShut {NoStop}%
\bibitem [{\citenamefont {Regazzoni}\ \emph {et~al.}(2021)\citenamefont
  {Regazzoni}, \citenamefont {Amri}, \citenamefont {Burri}, \citenamefont
  {Rusca}, \citenamefont {Zbinden},\ and\ \citenamefont
  {Charbon}}]{Regazzoni2021}%
  \BibitemOpen
  \bibfield  {author} {\bibinfo {author} {\bibfnamefont {F.}~\bibnamefont
  {Regazzoni}}, \bibinfo {author} {\bibfnamefont {E.}~\bibnamefont {Amri}},
  \bibinfo {author} {\bibfnamefont {S.}~\bibnamefont {Burri}}, \bibinfo
  {author} {\bibfnamefont {D.}~\bibnamefont {Rusca}}, \bibinfo {author}
  {\bibfnamefont {H.}~\bibnamefont {Zbinden}}, \ and\ \bibinfo {author}
  {\bibfnamefont {E.}~\bibnamefont {Charbon}},\ }\href@noop {} {\enquote
  {\bibinfo {title} {A high speed integrated quantum random number generator
  with on-chip real-time randomness extraction},}\ } (\bibinfo {year} {2021}),\
  \Eprint {http://arxiv.org/abs/2102.06238} {arXiv:2102.06238 [quant-ph]}
  \BibitemShut {NoStop}%
\bibitem [{\citenamefont {Thewes}\ \emph {et~al.}(2019)\citenamefont {Thewes},
  \citenamefont {L\"uders},\ and\ \citenamefont {A\ss{}mann}}]{QRNG_attack}%
  \BibitemOpen
  \bibfield  {author} {\bibinfo {author} {\bibfnamefont {J.}~\bibnamefont
  {Thewes}}, \bibinfo {author} {\bibfnamefont {C.}~\bibnamefont {L\"uders}}, \
  and\ \bibinfo {author} {\bibfnamefont {M.}~\bibnamefont {A\ss{}mann}},\
  }\href {\doibase 10.1103/PhysRevA.100.052318} {\bibfield  {journal} {\bibinfo
   {journal} {Phys. Rev. A}\ }\textbf {\bibinfo {volume} {100}},\ \bibinfo
  {pages} {052318} (\bibinfo {year} {2019})}\BibitemShut {NoStop}%
\bibitem [{\citenamefont {{Kuznetsov}}\ \emph {et~al.}(2019)\citenamefont
  {{Kuznetsov}}, \citenamefont {{Nariezhnii}}, \citenamefont {{Stelnyk}},
  \citenamefont {{Kokhanovska}}, \citenamefont {{Smirnov}},\ and\ \citenamefont
  {{Kuznetsova}}}]{QRNG_attack2}%
  \BibitemOpen
  \bibfield  {author} {\bibinfo {author} {\bibfnamefont {A.}~\bibnamefont
  {{Kuznetsov}}}, \bibinfo {author} {\bibfnamefont {O.}~\bibnamefont
  {{Nariezhnii}}}, \bibinfo {author} {\bibfnamefont {I.}~\bibnamefont
  {{Stelnyk}}}, \bibinfo {author} {\bibfnamefont {T.}~\bibnamefont
  {{Kokhanovska}}}, \bibinfo {author} {\bibfnamefont {O.}~\bibnamefont
  {{Smirnov}}}, \ and\ \bibinfo {author} {\bibfnamefont {T.}~\bibnamefont
  {{Kuznetsova}}},\ }\bibfield  {booktitle} {\emph {\bibinfo {booktitle} {2019
  10th IEEE International Conference on Intelligent Data Acquisition and
  Advanced Computing Systems: Technology and Applications (IDAACS)}},\ }\href
  {\doibase 10.1109/IDAACS.2019.8924447} {\ \textbf {\bibinfo {volume} {2}},\
  \bibinfo {pages} {713} (\bibinfo {year} {2019})}\BibitemShut {NoStop}%
\bibitem [{\citenamefont {Pironio}\ \emph {et~al.}(2010)\citenamefont
  {Pironio}, \citenamefont {Acín}, \citenamefont {Massar}, \citenamefont
  {de~la Giroday}, \citenamefont {Matsukevich}, \citenamefont {Maunz},
  \citenamefont {Olmschenk}, \citenamefont {Hayes}, \citenamefont {Luo},
  \citenamefont {Manning},\ and\ \citenamefont {et~al.}}]{DI_Pironio_2010}%
  \BibitemOpen
  \bibfield  {author} {\bibinfo {author} {\bibfnamefont {S.}~\bibnamefont
  {Pironio}}, \bibinfo {author} {\bibfnamefont {A.}~\bibnamefont {Acín}},
  \bibinfo {author} {\bibfnamefont {S.}~\bibnamefont {Massar}}, \bibinfo
  {author} {\bibfnamefont {A.~B.}\ \bibnamefont {de~la Giroday}}, \bibinfo
  {author} {\bibfnamefont {D.~N.}\ \bibnamefont {Matsukevich}}, \bibinfo
  {author} {\bibfnamefont {P.}~\bibnamefont {Maunz}}, \bibinfo {author}
  {\bibfnamefont {S.}~\bibnamefont {Olmschenk}}, \bibinfo {author}
  {\bibfnamefont {D.}~\bibnamefont {Hayes}}, \bibinfo {author} {\bibfnamefont
  {L.}~\bibnamefont {Luo}}, \bibinfo {author} {\bibfnamefont {T.~A.}\
  \bibnamefont {Manning}}, \ and\ \bibinfo {author} {\bibnamefont {et~al.}},\
  }\href {\doibase 10.1038/nature09008} {\bibfield  {journal} {\bibinfo
  {journal} {Nature}\ }\textbf {\bibinfo {volume} {464}},\ \bibinfo {pages}
  {1021–1024} (\bibinfo {year} {2010})}\BibitemShut {NoStop}%
\bibitem [{\citenamefont {{Brown}}\ \emph {et~al.}(2020)\citenamefont
  {{Brown}}, \citenamefont {{Ragy}},\ and\ \citenamefont
  {{Colbeck}}}]{DI_roger}%
  \BibitemOpen
  \bibfield  {author} {\bibinfo {author} {\bibfnamefont {P.~J.}\ \bibnamefont
  {{Brown}}}, \bibinfo {author} {\bibfnamefont {S.}~\bibnamefont {{Ragy}}}, \
  and\ \bibinfo {author} {\bibfnamefont {R.}~\bibnamefont {{Colbeck}}},\ }\href
  {\doibase 10.1109/TIT.2019.2960252} {\bibfield  {journal} {\bibinfo
  {journal} {IEEE Transactions on Information Theory}\ }\textbf {\bibinfo
  {volume} {66}},\ \bibinfo {pages} {2964} (\bibinfo {year}
  {2020})}\BibitemShut {NoStop}%
\bibitem [{\citenamefont {Liu}\ \emph {et~al.}(2021)\citenamefont {Liu} \emph
  {et~al.}}]{Liu2021}%
  \BibitemOpen
  \bibfield  {author} {\bibinfo {author} {\bibfnamefont {W.-Z.}\ \bibnamefont
  {Liu}} \emph {et~al.},\ }\href {\doibase 10.1038/s41567-020-01147-2}
  {\bibfield  {journal} {\bibinfo  {journal} {Nature Physics}\ } (\bibinfo
  {year} {2021}),\ 10.1038/s41567-020-01147-2}\BibitemShut {NoStop}%
\bibitem [{\citenamefont {Liu}\ \emph {et~al.}(2018)\citenamefont {Liu} \emph
  {et~al.}}]{Liu2018a}%
  \BibitemOpen
  \bibfield  {author} {\bibinfo {author} {\bibfnamefont {Y.}~\bibnamefont
  {Liu}} \emph {et~al.},\ }\href {\doibase 10.1038/s41586-018-0559-3}
  {\bibfield  {journal} {\bibinfo  {journal} {Nature}\ }\textbf {\bibinfo
  {volume} {562}},\ \bibinfo {pages} {548} (\bibinfo {year} {2018})},\ \Eprint
  {http://arxiv.org/abs/1807.09611} {arXiv:1807.09611} \BibitemShut {NoStop}%
\bibitem [{\citenamefont {Zhang}\ \emph {et~al.}(2020)\citenamefont {Zhang}
  \emph {et~al.}}]{Zhang2020}%
  \BibitemOpen
  \bibfield  {author} {\bibinfo {author} {\bibfnamefont {Y.}~\bibnamefont
  {Zhang}} \emph {et~al.},\ }\href {\doibase 10.1103/PhysRevLett.124.010505}
  {\bibfield  {journal} {\bibinfo  {journal} {Phys. Rev. Lett.}\ }\textbf
  {\bibinfo {volume} {124}},\ \bibinfo {pages} {010505} (\bibinfo {year}
  {2020})}\BibitemShut {NoStop}%
\bibitem [{\citenamefont {Li}\ \emph {et~al.}(2021)\citenamefont {Li} \emph
  {et~al.}}]{DI_new}%
  \BibitemOpen
  \bibfield  {author} {\bibinfo {author} {\bibfnamefont {M.-H.}\ \bibnamefont
  {Li}} \emph {et~al.},\ }\href {\doibase 10.1103/PhysRevLett.126.050503}
  {\bibfield  {journal} {\bibinfo  {journal} {Phys. Rev. Lett.}\ }\textbf
  {\bibinfo {volume} {126}},\ \bibinfo {pages} {050503} (\bibinfo {year}
  {2021})}\BibitemShut {NoStop}%
\bibitem [{\citenamefont {Foletto}\ \emph {et~al.}(2021)\citenamefont
  {Foletto}, \citenamefont {Padovan}, \citenamefont {Avesani}, \citenamefont
  {Tebyanian}, \citenamefont {Villoresi},\ and\ \citenamefont
  {Vallone}}]{foletto2021experimental}%
  \BibitemOpen
  \bibfield  {author} {\bibinfo {author} {\bibfnamefont {G.}~\bibnamefont
  {Foletto}}, \bibinfo {author} {\bibfnamefont {M.}~\bibnamefont {Padovan}},
  \bibinfo {author} {\bibfnamefont {M.}~\bibnamefont {Avesani}}, \bibinfo
  {author} {\bibfnamefont {H.}~\bibnamefont {Tebyanian}}, \bibinfo {author}
  {\bibfnamefont {P.}~\bibnamefont {Villoresi}}, \ and\ \bibinfo {author}
  {\bibfnamefont {G.}~\bibnamefont {Vallone}},\ }\href@noop {} {\enquote
  {\bibinfo {title} {Experimental test of sequential weak measurements for
  certified quantum randomness extraction},}\ } (\bibinfo {year} {2021}),\
  \Eprint {http://arxiv.org/abs/2101.12074} {arXiv:2101.12074 [quant-ph]}
  \BibitemShut {NoStop}%
\bibitem [{\citenamefont {{\v{S}}upi{\'{c}}}\ and\ \citenamefont
  {Bowles}(2020)}]{Supic2020}%
  \BibitemOpen
  \bibfield  {author} {\bibinfo {author} {\bibfnamefont {I.}~\bibnamefont
  {{\v{S}}upi{\'{c}}}}\ and\ \bibinfo {author} {\bibfnamefont {J.}~\bibnamefont
  {Bowles}},\ }\href {\doibase 10.22331/q-2020-09-30-337} {\bibfield  {journal}
  {\bibinfo  {journal} {{Quantum}}\ }\textbf {\bibinfo {volume} {4}},\ \bibinfo
  {pages} {337} (\bibinfo {year} {2020})}\BibitemShut {NoStop}%
\bibitem [{\citenamefont {Tavakoli}(2021)}]{tavakoli2021semideviceindependent}%
  \BibitemOpen
  \bibfield  {author} {\bibinfo {author} {\bibfnamefont {A.}~\bibnamefont
  {Tavakoli}},\ }\href@noop {} {\enquote {\bibinfo {title}
  {Semi-device-independent framework based on restricted distrust in
  prepare-and-measure experiments},}\ } (\bibinfo {year} {2021}),\ \Eprint
  {http://arxiv.org/abs/2101.07830} {arXiv:2101.07830 [quant-ph]} \BibitemShut
  {NoStop}%
\bibitem [{\citenamefont {Cao}\ \emph {et~al.}(2016)\citenamefont {Cao},
  \citenamefont {Zhou}, \citenamefont {Yuan},\ and\ \citenamefont
  {Ma}}]{Ma_new1}%
  \BibitemOpen
  \bibfield  {author} {\bibinfo {author} {\bibfnamefont {Z.}~\bibnamefont
  {Cao}}, \bibinfo {author} {\bibfnamefont {H.}~\bibnamefont {Zhou}}, \bibinfo
  {author} {\bibfnamefont {X.}~\bibnamefont {Yuan}}, \ and\ \bibinfo {author}
  {\bibfnamefont {X.}~\bibnamefont {Ma}},\ }\href {\doibase
  10.1103/PhysRevX.6.011020} {\bibfield  {journal} {\bibinfo  {journal} {Phys.
  Rev. X}\ }\textbf {\bibinfo {volume} {6}},\ \bibinfo {pages} {011020}
  (\bibinfo {year} {2016})}\BibitemShut {NoStop}%
\bibitem [{\citenamefont {Avesani}\ \emph {et~al.}(2020)\citenamefont
  {Avesani}, \citenamefont {Tebyanian}, \citenamefont {Villoresi},\ and\
  \citenamefont {Vallone}}]{avesani2020unbounded}%
  \BibitemOpen
  \bibfield  {author} {\bibinfo {author} {\bibfnamefont {M.}~\bibnamefont
  {Avesani}}, \bibinfo {author} {\bibfnamefont {H.}~\bibnamefont {Tebyanian}},
  \bibinfo {author} {\bibfnamefont {P.}~\bibnamefont {Villoresi}}, \ and\
  \bibinfo {author} {\bibfnamefont {G.}~\bibnamefont {Vallone}},\ }\href@noop
  {} {\enquote {\bibinfo {title} {Unbounded randomness from uncharacterized
  sources},}\ } (\bibinfo {year} {2020}),\ \Eprint
  {http://arxiv.org/abs/2010.05798} {arXiv:2010.05798 [quant-ph]} \BibitemShut
  {NoStop}%
\bibitem [{\citenamefont {Avesani}\ \emph {et~al.}(2018)\citenamefont
  {Avesani}, \citenamefont {Marangon}, \citenamefont {Vallone},\ and\
  \citenamefont {Villoresi}}]{Avesani2018}%
  \BibitemOpen
  \bibfield  {author} {\bibinfo {author} {\bibfnamefont {M.}~\bibnamefont
  {Avesani}}, \bibinfo {author} {\bibfnamefont {D.~G.}\ \bibnamefont
  {Marangon}}, \bibinfo {author} {\bibfnamefont {G.}~\bibnamefont {Vallone}}, \
  and\ \bibinfo {author} {\bibfnamefont {P.}~\bibnamefont {Villoresi}},\ }\href
  {\doibase 10.1038/s41467-018-07585-0} {\bibfield  {journal} {\bibinfo
  {journal} {Nature Communications}\ }\textbf {\bibinfo {volume} {9}},\
  \bibinfo {pages} {5365} (\bibinfo {year} {2018})}\BibitemShut {NoStop}%
\bibitem [{\citenamefont {Drahi}\ \emph {et~al.}(2020)\citenamefont {Drahi},
  \citenamefont {Walk}, \citenamefont {Hoban}, \citenamefont {Fedorov},
  \citenamefont {Shakhovoy}, \citenamefont {Feimov}, \citenamefont {Kurochkin},
  \citenamefont {Kolthammer}, \citenamefont {Nunn}, \citenamefont {Barrett},\
  and\ \citenamefont {Walmsley}}]{PhysRevX_SDI}%
  \BibitemOpen
  \bibfield  {author} {\bibinfo {author} {\bibfnamefont {D.}~\bibnamefont
  {Drahi}}, \bibinfo {author} {\bibfnamefont {N.}~\bibnamefont {Walk}},
  \bibinfo {author} {\bibfnamefont {M.~J.}\ \bibnamefont {Hoban}}, \bibinfo
  {author} {\bibfnamefont {A.~K.}\ \bibnamefont {Fedorov}}, \bibinfo {author}
  {\bibfnamefont {R.}~\bibnamefont {Shakhovoy}}, \bibinfo {author}
  {\bibfnamefont {A.}~\bibnamefont {Feimov}}, \bibinfo {author} {\bibfnamefont
  {Y.}~\bibnamefont {Kurochkin}}, \bibinfo {author} {\bibfnamefont {W.~S.}\
  \bibnamefont {Kolthammer}}, \bibinfo {author} {\bibfnamefont
  {J.}~\bibnamefont {Nunn}}, \bibinfo {author} {\bibfnamefont {J.}~\bibnamefont
  {Barrett}}, \ and\ \bibinfo {author} {\bibfnamefont {I.~A.}\ \bibnamefont
  {Walmsley}},\ }\href {\doibase 10.1103/PhysRevX.10.041048} {\bibfield
  {journal} {\bibinfo  {journal} {Phys. Rev. X}\ }\textbf {\bibinfo {volume}
  {10}},\ \bibinfo {pages} {041048} (\bibinfo {year} {2020})}\BibitemShut
  {NoStop}%
\bibitem [{\citenamefont {Cao}\ \emph {et~al.}(2015)\citenamefont {Cao},
  \citenamefont {Zhou},\ and\ \citenamefont {Ma}}]{MDI}%
  \BibitemOpen
  \bibfield  {author} {\bibinfo {author} {\bibfnamefont {Z.}~\bibnamefont
  {Cao}}, \bibinfo {author} {\bibfnamefont {H.}~\bibnamefont {Zhou}}, \ and\
  \bibinfo {author} {\bibfnamefont {X.}~\bibnamefont {Ma}},\ }\href {\doibase
  10.1088/1367-2630/17/12/125011} {\bibfield  {journal} {\bibinfo  {journal}
  {New Journal of Physics}\ }\textbf {\bibinfo {volume} {17}},\ \bibinfo
  {pages} {125011} (\bibinfo {year} {2015})}\BibitemShut {NoStop}%
\bibitem [{\citenamefont {Nie}\ \emph {et~al.}(2016)\citenamefont {Nie},
  \citenamefont {Guan}, \citenamefont {Zhou}, \citenamefont {Zhang},
  \citenamefont {Ma}, \citenamefont {Zhang},\ and\ \citenamefont {Pan}}]{MDI2}%
  \BibitemOpen
  \bibfield  {author} {\bibinfo {author} {\bibfnamefont {Y.-Q.}\ \bibnamefont
  {Nie}}, \bibinfo {author} {\bibfnamefont {J.-Y.}\ \bibnamefont {Guan}},
  \bibinfo {author} {\bibfnamefont {H.}~\bibnamefont {Zhou}}, \bibinfo {author}
  {\bibfnamefont {Q.}~\bibnamefont {Zhang}}, \bibinfo {author} {\bibfnamefont
  {X.}~\bibnamefont {Ma}}, \bibinfo {author} {\bibfnamefont {J.}~\bibnamefont
  {Zhang}}, \ and\ \bibinfo {author} {\bibfnamefont {J.-W.}\ \bibnamefont
  {Pan}},\ }\href {\doibase 10.1103/physreva.94.060301} {\bibfield  {journal}
  {\bibinfo  {journal} {Physical Review A}\ }\textbf {\bibinfo {volume} {94}}
  (\bibinfo {year} {2016}),\ 10.1103/physreva.94.060301}\BibitemShut {NoStop}%
\bibitem [{\citenamefont {Tebyanian}\ \emph {et~al.}(2020)\citenamefont
  {Tebyanian}, \citenamefont {Avesani}, \citenamefont {Vallone},\ and\
  \citenamefont {Villoresi}}]{tebyanian2020semidevice}%
  \BibitemOpen
  \bibfield  {author} {\bibinfo {author} {\bibfnamefont {H.}~\bibnamefont
  {Tebyanian}}, \bibinfo {author} {\bibfnamefont {M.}~\bibnamefont {Avesani}},
  \bibinfo {author} {\bibfnamefont {G.}~\bibnamefont {Vallone}}, \ and\
  \bibinfo {author} {\bibfnamefont {P.}~\bibnamefont {Villoresi}},\ }\href@noop
  {} {\enquote {\bibinfo {title} {Semi-device independent randomness from
  d-outcome continuous-variable detection},}\ } (\bibinfo {year} {2020}),\
  \Eprint {http://arxiv.org/abs/2009.08897} {arXiv:2009.08897 [quant-ph]}
  \BibitemShut {NoStop}%
\bibitem [{\citenamefont {Brask}\ \emph {et~al.}(2017)\citenamefont {Brask},
  \citenamefont {Martin}, \citenamefont {Esposito}, \citenamefont {Houlmann},
  \citenamefont {Bowles}, \citenamefont {Zbinden},\ and\ \citenamefont
  {Brunner}}]{Brask2017}%
  \BibitemOpen
  \bibfield  {author} {\bibinfo {author} {\bibfnamefont {J.~B.}\ \bibnamefont
  {Brask}}, \bibinfo {author} {\bibfnamefont {A.}~\bibnamefont {Martin}},
  \bibinfo {author} {\bibfnamefont {W.}~\bibnamefont {Esposito}}, \bibinfo
  {author} {\bibfnamefont {R.}~\bibnamefont {Houlmann}}, \bibinfo {author}
  {\bibfnamefont {J.}~\bibnamefont {Bowles}}, \bibinfo {author} {\bibfnamefont
  {H.}~\bibnamefont {Zbinden}}, \ and\ \bibinfo {author} {\bibfnamefont
  {N.}~\bibnamefont {Brunner}},\ }\href {\doibase
  10.1103/PhysRevApplied.7.054018} {\bibfield  {journal} {\bibinfo  {journal}
  {Physical Review Applied}\ }\textbf {\bibinfo {volume} {7}},\ \bibinfo
  {pages} {054018} (\bibinfo {year} {2017})}\BibitemShut {NoStop}%
\bibitem [{\citenamefont {Rusca}\ \emph {et~al.}(2020)\citenamefont {Rusca},
  \citenamefont {Tebyanian}, \citenamefont {Martin},\ and\ \citenamefont
  {Zbinden}}]{Rusca2020}%
  \BibitemOpen
  \bibfield  {author} {\bibinfo {author} {\bibfnamefont {D.}~\bibnamefont
  {Rusca}}, \bibinfo {author} {\bibfnamefont {H.}~\bibnamefont {Tebyanian}},
  \bibinfo {author} {\bibfnamefont {A.}~\bibnamefont {Martin}}, \ and\ \bibinfo
  {author} {\bibfnamefont {H.}~\bibnamefont {Zbinden}},\ }\href {\doibase
  10.1063/5.0011479} {\bibfield  {journal} {\bibinfo  {journal} {Applied
  Physics Letters}\ }\textbf {\bibinfo {volume} {116}} (\bibinfo {year}
  {2020}),\ 10.1063/5.0011479},\ \Eprint {http://arxiv.org/abs/2004.08307}
  {arXiv:2004.08307} \BibitemShut {NoStop}%
\bibitem [{\citenamefont {Avesani}\ \emph {et~al.}(2021)\citenamefont
  {Avesani}, \citenamefont {Tebyanian}, \citenamefont {Villoresi},\ and\
  \citenamefont {Vallone}}]{avesani2020}%
  \BibitemOpen
  \bibfield  {author} {\bibinfo {author} {\bibfnamefont {M.}~\bibnamefont
  {Avesani}}, \bibinfo {author} {\bibfnamefont {H.}~\bibnamefont {Tebyanian}},
  \bibinfo {author} {\bibfnamefont {P.}~\bibnamefont {Villoresi}}, \ and\
  \bibinfo {author} {\bibfnamefont {G.}~\bibnamefont {Vallone}},\ }\href
  {\doibase 10.1103/PhysRevApplied.15.034034} {\bibfield  {journal} {\bibinfo
  {journal} {Phys. Rev. Applied}\ }\textbf {\bibinfo {volume} {15}},\ \bibinfo
  {pages} {034034} (\bibinfo {year} {2021})}\BibitemShut {NoStop}%
\bibitem [{\citenamefont {Barnett}\ and\ \citenamefont
  {Croke}(2009)}]{state_dis}%
  \BibitemOpen
  \bibfield  {author} {\bibinfo {author} {\bibfnamefont {S.~M.}\ \bibnamefont
  {Barnett}}\ and\ \bibinfo {author} {\bibfnamefont {S.}~\bibnamefont
  {Croke}},\ }\href {\doibase 10.1364/AOP.1.000238} {\bibfield  {journal}
  {\bibinfo  {journal} {Adv. Opt. Photon.}\ }\textbf {\bibinfo {volume} {1}},\
  \bibinfo {pages} {238} (\bibinfo {year} {2009})}\BibitemShut {NoStop}%
\bibitem [{\citenamefont {Van~Himbeeck}\ \emph {et~al.}(2017)\citenamefont
  {Van~Himbeeck}, \citenamefont {Woodhead}, \citenamefont {Cerf}, \citenamefont
  {Garc{\'{i}}a-Patr{\'{o}}n},\ and\ \citenamefont
  {Pironio}}]{VanHimbeeck2017semidevice}%
  \BibitemOpen
  \bibfield  {author} {\bibinfo {author} {\bibfnamefont {T.}~\bibnamefont
  {Van~Himbeeck}}, \bibinfo {author} {\bibfnamefont {E.}~\bibnamefont
  {Woodhead}}, \bibinfo {author} {\bibfnamefont {N.~J.}\ \bibnamefont {Cerf}},
  \bibinfo {author} {\bibfnamefont {R.}~\bibnamefont
  {Garc{\'{i}}a-Patr{\'{o}}n}}, \ and\ \bibinfo {author} {\bibfnamefont
  {S.}~\bibnamefont {Pironio}},\ }\href {\doibase 10.22331/q-2017-11-18-33}
  {\bibfield  {journal} {\bibinfo  {journal} {{Quantum}}\ }\textbf {\bibinfo
  {volume} {1}},\ \bibinfo {pages} {33} (\bibinfo {year} {2017})}\BibitemShut
  {NoStop}%
\bibitem [{\citenamefont {Gras}\ \emph {et~al.}(2020)\citenamefont {Gras},
  \citenamefont {Martin}, \citenamefont {Choi},\ and\ \citenamefont
  {Bussières}}]{gras2020quantum}%
  \BibitemOpen
  \bibfield  {author} {\bibinfo {author} {\bibfnamefont {G.}~\bibnamefont
  {Gras}}, \bibinfo {author} {\bibfnamefont {A.}~\bibnamefont {Martin}},
  \bibinfo {author} {\bibfnamefont {J.~W.}\ \bibnamefont {Choi}}, \ and\
  \bibinfo {author} {\bibfnamefont {F.}~\bibnamefont {Bussières}},\
  }\href@noop {} {\enquote {\bibinfo {title} {Quantum entropy model of an
  integrated qrng chip},}\ } (\bibinfo {year} {2020}),\ \Eprint
  {http://arxiv.org/abs/2011.14129} {arXiv:2011.14129 [quant-ph]} \BibitemShut
  {NoStop}%
\bibitem [{\citenamefont {Leone}\ \emph {et~al.}(2020)\citenamefont {Leone},
  \citenamefont {Rusca}, \citenamefont {Azzini}, \citenamefont {Fontana},
  \citenamefont {Acerbi}, \citenamefont {Gola}, \citenamefont {Tontini},
  \citenamefont {Massari}, \citenamefont {Zbinden},\ and\ \citenamefont
  {Pavesi}}]{QRNG_chip}%
  \BibitemOpen
  \bibfield  {author} {\bibinfo {author} {\bibfnamefont {N.}~\bibnamefont
  {Leone}}, \bibinfo {author} {\bibfnamefont {D.}~\bibnamefont {Rusca}},
  \bibinfo {author} {\bibfnamefont {S.}~\bibnamefont {Azzini}}, \bibinfo
  {author} {\bibfnamefont {G.}~\bibnamefont {Fontana}}, \bibinfo {author}
  {\bibfnamefont {F.}~\bibnamefont {Acerbi}}, \bibinfo {author} {\bibfnamefont
  {A.}~\bibnamefont {Gola}}, \bibinfo {author} {\bibfnamefont {A.}~\bibnamefont
  {Tontini}}, \bibinfo {author} {\bibfnamefont {N.}~\bibnamefont {Massari}},
  \bibinfo {author} {\bibfnamefont {H.}~\bibnamefont {Zbinden}}, \ and\
  \bibinfo {author} {\bibfnamefont {L.}~\bibnamefont {Pavesi}},\ }\href
  {\doibase 10.1063/5.0022526} {\bibfield  {journal} {\bibinfo  {journal} {APL
  Photonics}\ }\textbf {\bibinfo {volume} {5}},\ \bibinfo {pages} {101301}
  (\bibinfo {year} {2020})},\ \Eprint
  {http://arxiv.org/abs/https://doi.org/10.1063/5.0022526}
  {https://doi.org/10.1063/5.0022526} \BibitemShut {NoStop}%
\bibitem [{\citenamefont {Van~Himbeeck}\ and\ \citenamefont
  {Pironio}(2019)}]{Himbeeck2019CorrelationsConstraints}%
  \BibitemOpen
  \bibfield  {author} {\bibinfo {author} {\bibfnamefont {T.}~\bibnamefont
  {Van~Himbeeck}}\ and\ \bibinfo {author} {\bibfnamefont {S.}~\bibnamefont
  {Pironio}},\ }\href {https://arxiv.org/abs/1905.09117} {\bibfield  {journal}
  {\bibinfo  {journal} {arXiv preprint arXiv:1905.09117}\ } (\bibinfo {year}
  {2019})}\BibitemShut {NoStop}%
\bibitem [{\citenamefont {{Konig}}\ \emph {et~al.}(2009)\citenamefont
  {{Konig}}, \citenamefont {{Renner}},\ and\ \citenamefont
  {{Schaffner}}}]{entropy}%
  \BibitemOpen
  \bibfield  {author} {\bibinfo {author} {\bibfnamefont {R.}~\bibnamefont
  {{Konig}}}, \bibinfo {author} {\bibfnamefont {R.}~\bibnamefont {{Renner}}}, \
  and\ \bibinfo {author} {\bibfnamefont {C.}~\bibnamefont {{Schaffner}}},\
  }\href {\doibase 10.1109/TIT.2009.2025545} {\bibfield  {journal} {\bibinfo
  {journal} {IEEE Transactions on Information Theory}\ }\textbf {\bibinfo
  {volume} {55}},\ \bibinfo {pages} {4337} (\bibinfo {year}
  {2009})}\BibitemShut {NoStop}%
\bibitem [{\citenamefont {Tomamichel}\ \emph {et~al.}(2011)\citenamefont
  {Tomamichel}, \citenamefont {Schaffner}, \citenamefont {Smith},\ and\
  \citenamefont {Renner}}]{Tomamichel2011}%
  \BibitemOpen
  \bibfield  {author} {\bibinfo {author} {\bibfnamefont {M.}~\bibnamefont
  {Tomamichel}}, \bibinfo {author} {\bibfnamefont {C.}~\bibnamefont
  {Schaffner}}, \bibinfo {author} {\bibfnamefont {A.}~\bibnamefont {Smith}}, \
  and\ \bibinfo {author} {\bibfnamefont {R.}~\bibnamefont {Renner}},\ }\href
  {\doibase 10.1109/TIT.2011.2158473} {\bibfield  {journal} {\bibinfo
  {journal} {IEEE Transactions on Information Theory}\ }\textbf {\bibinfo
  {volume} {57}},\ \bibinfo {pages} {5524} (\bibinfo {year}
  {2011})}\BibitemShut {NoStop}%
\bibitem [{\citenamefont {Caloz}\ \emph {et~al.}(2018)\citenamefont {Caloz},
  \citenamefont {Perrenoud}, \citenamefont {Autebert}, \citenamefont {Korzh},
  \citenamefont {Weiss}, \citenamefont {Schönenberger}, \citenamefont
  {Warburton}, \citenamefont {Zbinden},\ and\ \citenamefont
  {Bussières}}]{SNSPD}%
  \BibitemOpen
  \bibfield  {author} {\bibinfo {author} {\bibfnamefont {M.}~\bibnamefont
  {Caloz}}, \bibinfo {author} {\bibfnamefont {M.}~\bibnamefont {Perrenoud}},
  \bibinfo {author} {\bibfnamefont {C.}~\bibnamefont {Autebert}}, \bibinfo
  {author} {\bibfnamefont {B.}~\bibnamefont {Korzh}}, \bibinfo {author}
  {\bibfnamefont {M.}~\bibnamefont {Weiss}}, \bibinfo {author} {\bibfnamefont
  {C.}~\bibnamefont {Schönenberger}}, \bibinfo {author} {\bibfnamefont
  {R.~J.}\ \bibnamefont {Warburton}}, \bibinfo {author} {\bibfnamefont
  {H.}~\bibnamefont {Zbinden}}, \ and\ \bibinfo {author} {\bibfnamefont
  {F.}~\bibnamefont {Bussières}},\ }\href {\doibase 10.1063/1.5010102}
  {\bibfield  {journal} {\bibinfo  {journal} {Applied Physics Letters}\
  }\textbf {\bibinfo {volume} {112}},\ \bibinfo {pages} {061103} (\bibinfo
  {year} {2018})},\ \Eprint
  {http://arxiv.org/abs/https://doi.org/10.1063/1.5010102}
  {https://doi.org/10.1063/1.5010102} \BibitemShut {NoStop}%
\bibitem [{\citenamefont {Eisaman}\ \emph {et~al.}(2011)\citenamefont
  {Eisaman}, \citenamefont {Fan}, \citenamefont {Migdall},\ and\ \citenamefont
  {Polyakov}}]{SPD}%
  \BibitemOpen
  \bibfield  {author} {\bibinfo {author} {\bibfnamefont {M.~D.}\ \bibnamefont
  {Eisaman}}, \bibinfo {author} {\bibfnamefont {J.}~\bibnamefont {Fan}},
  \bibinfo {author} {\bibfnamefont {A.}~\bibnamefont {Migdall}}, \ and\
  \bibinfo {author} {\bibfnamefont {S.~V.}\ \bibnamefont {Polyakov}},\ }\href
  {\doibase 10.1063/1.3610677} {\bibfield  {journal} {\bibinfo  {journal}
  {Review of Scientific Instruments}\ }\textbf {\bibinfo {volume} {82}},\
  \bibinfo {pages} {071101} (\bibinfo {year} {2011})},\ \Eprint
  {http://arxiv.org/abs/https://doi.org/10.1063/1.3610677}
  {https://doi.org/10.1063/1.3610677} \BibitemShut {NoStop}%
\bibitem [{\citenamefont {Roberts}\ \emph {et~al.}(2018)\citenamefont
  {Roberts}, \citenamefont {Pittaluga}, \citenamefont {Minder}, \citenamefont
  {Lucamarini}, \citenamefont {Dynes}, \citenamefont {Yuan},\ and\
  \citenamefont {Shields}}]{Roberts:18}%
  \BibitemOpen
  \bibfield  {author} {\bibinfo {author} {\bibfnamefont {G.~L.}\ \bibnamefont
  {Roberts}}, \bibinfo {author} {\bibfnamefont {M.}~\bibnamefont {Pittaluga}},
  \bibinfo {author} {\bibfnamefont {M.}~\bibnamefont {Minder}}, \bibinfo
  {author} {\bibfnamefont {M.}~\bibnamefont {Lucamarini}}, \bibinfo {author}
  {\bibfnamefont {J.~F.}\ \bibnamefont {Dynes}}, \bibinfo {author}
  {\bibfnamefont {Z.~L.}\ \bibnamefont {Yuan}}, \ and\ \bibinfo {author}
  {\bibfnamefont {A.~J.}\ \bibnamefont {Shields}},\ }\href {\doibase
  10.1364/OL.43.005110} {\bibfield  {journal} {\bibinfo  {journal} {Opt.
  Lett.}\ }\textbf {\bibinfo {volume} {43}},\ \bibinfo {pages} {5110} (\bibinfo
  {year} {2018})}\BibitemShut {NoStop}%
\bibitem [{\citenamefont {Bancal}\ \emph {et~al.}(2014)\citenamefont {Bancal},
  \citenamefont {Sheridan},\ and\ \citenamefont {Scarani}}]{Bancal14}%
  \BibitemOpen
  \bibfield  {author} {\bibinfo {author} {\bibfnamefont {J.-D.}\ \bibnamefont
  {Bancal}}, \bibinfo {author} {\bibfnamefont {L.}~\bibnamefont {Sheridan}}, \
  and\ \bibinfo {author} {\bibfnamefont {V.}~\bibnamefont {Scarani}},\ }\href
  {\doibase 10.1088/1367-2630/16/3/033011} {\bibfield  {journal} {\bibinfo
  {journal} {New Journal of Physics}\ }\textbf {\bibinfo {volume} {16}},\
  \bibinfo {pages} {033011} (\bibinfo {year} {2014})}\BibitemShut {NoStop}%
\bibitem [{\citenamefont {Boyd}\ \emph {et~al.}(2004)\citenamefont {Boyd},
  \citenamefont {Boyd},\ and\ \citenamefont {Vandenberghe}}]{boyd2004convex}%
  \BibitemOpen
  \bibfield  {author} {\bibinfo {author} {\bibfnamefont {S.}~\bibnamefont
  {Boyd}}, \bibinfo {author} {\bibfnamefont {S.~P.}\ \bibnamefont {Boyd}}, \
  and\ \bibinfo {author} {\bibfnamefont {L.}~\bibnamefont {Vandenberghe}},\
  }\href@noop {} {\emph {\bibinfo {title} {Convex optimization}}}\ (\bibinfo
  {publisher} {Cambridge university press},\ \bibinfo {year}
  {2004})\BibitemShut {NoStop}%
\end{thebibliography}%
\appendix
\section{Generalized Semi-definite Programming for $n$-input $d$-outcome}
\label{app:a}
\subsection{Primal}
This appendix 
{presents a generalized expression }of 
the guessing probability optimization problem shown in Eq.\ref{eq:P_g}, in the form of a semidefinite program (SDP). This optimization is used to derive a bound on the min-entropy for a $n$-input $d$-outcome semi-DI QRNG protocol based on an energy bound,
generalizing the approach proposed in~\cite{Brask2017}.
The generalized form of guessing probability for $n$-input $d$-outcomes reads:

\beq
\begin{aligned}
P_{\rm guess} =
\frac{1}{n}  \max\limits_{\{\rho^{\boldsymbol\lambda}_x, p_{\boldsymbol\lambda} ,\Pi _b^{\boldsymbol\lambda}\}}
\left( 
\sum\limits_{x = 0}^{n-1} \sum\limits_{\boldsymbol\lambda} {p_{\boldsymbol\lambda}} \max_b  \bigg\{ \Tr[\rho^{\boldsymbol\lambda}_x \Pi^{\boldsymbol\lambda}_b] \bigg\} \right)\,,
\end{aligned}
\label{eq:P_g33}
\eeq
where $\Pi^{\boldsymbol\lambda}_b$ with $b=0,\cdots,d-1$ represent
positive-operator valued measurement (POVM) operators 
in a $n$ dimensional Hilbert space
and the states $\rho_x$ satisfy
the constraint
$p(b|x) = \sum\limits_{\boldsymbol\lambda}  {p_{\boldsymbol\lambda}} 
\tr[\hat\rho_x \hat\Pi_b^{\boldsymbol\lambda}]
$.
In the above equation, we assume the probability of transmitting $x \in \{0,\dots,n-1\}$ is identical and equal to $p_{x} = \frac{1}{n} $.
The variable $\boldsymbol\lambda$ labels a possible ``strategy''.
As discussed in \cite{Brask2017} and \cite{Bancal14}, all strategies 
in which the inner maximization over $b$ in equation (\ref{eq:P_g33}) occurs for the same value of $b$ at given $x$ can be grouped.
Consequently, it is sufficient to consider at most $d^{n}$ strategies when maximizing equation (\ref{eq:P_g33}) over all potential measurement strategies. Then, each strategy 
can be labeled as 
${\Lambda}=(\lambda_0,\cdots,\lambda_{n-1})$, where $\lambda_k=0,\cdots, d-1$, and $\sum_{\Lambda } := \sum_{\lambda_0 = 0}^{d-1}\cdots \sum_{\lambda_{n-1} = 0}^{d-1}$ is defined for simplicity. The value of $\lambda_x$ indicates that the $b=\lambda_x$ outcome maximizes  
$\Tr[\rho_x \Pi^{\Lambda}_b]$ when the state $\hat\rho_x$ is sent.
By absorbing the weight $p_{\Lambda}$ into the normalization of POVMs, 
$M _b^{\Lambda}=p_{\Lambda} \Pi _b^{\Lambda}$, Eq. (\ref{eq:P_g33}) can be rewritten as
\beq
{P_{\rm guess}} = \frac{1}{n}\mathop{\max }\limits_{\{M _b^{\Lambda}\}} \sum\limits_{x = 0}^{n-1}
{\sum\limits_{\Lambda} \Tr[\rho_x M _{\lambda_x}^{\Lambda} ]},
\label{eq:P_g3}
\eeq
As discussed in the main text, the states $\rho_x$ can
be chosen to be pure $\rho_x=\ket{\psi_x}\bra{\psi_x}$. If the energy constraint is imposed, then the states $\{\ket{\psi_0}, \ket{\psi_1}, \dots, \ket{\psi_{n-1}}\}$ can be 
{can be express as a linear combination of an orthonormal basis} $\{\ket{0},\ket{1}, \dots, \ket{n-1}\}$ 
with
fixed overlap $|\braket{\psi_x}{\psi_y}|=\delta$ for $x\neq y$.

The maximization of the guessing probability $P_{\rm guess}$ can be cast as an SDP,  
whose primal form can be written as follows
\begin{maxi}|l|
  {
  M_b^{\Lambda}
  }
  { P_{\rm g}=
  \frac{1}{n}\sum\limits_{x = 0}^{n-1} 
  \sum\limits_{\Lambda} \bra{\psi_x}
  M_{\lambda_x}^{\Lambda}\ket{\psi_x}
  }
  {}{}
  \addConstraint{M_b^{\Lambda} = (M_b^{\Lambda})^\dag }
  \addConstraint{M_b^{\Lambda} \ge 0}
  \addConstraint{\sum_{b=0}^{d-1} {M_b^{\Lambda}  = \frac1n \tr[\sum_{b=0}^{d-1} M_b^{\Lambda}}]\mathbb{I}}
  \addConstraint{\sum\limits_{\Lambda}
  \bra{\psi_x}
  M_{b}^{\Lambda}
  \ket{\psi_x}
  = p(b|x)\,,\quad \forall b,x}
  \label{eq:SDP_primal}
\end{maxi}
where $M_{b}^{\Lambda}$ are $n\times n$ Hermitian semi-positive matrices.
This maximization defines an SDP, converging to optimal bounds on $P_{\rm guess}$ 
given the constraints on the overlap or the energy and the observed data $p(b|x)$.

The maximization is performed over all measurement strategies $M _b^{\Lambda}$ meaning that the computational cost increases with the number of outcomes. 
In this case, we can also derive the dual SDP, whose derivation is described in the next section.
\begin{figure}[h!]
\centering
\includegraphics[width=\linewidth]{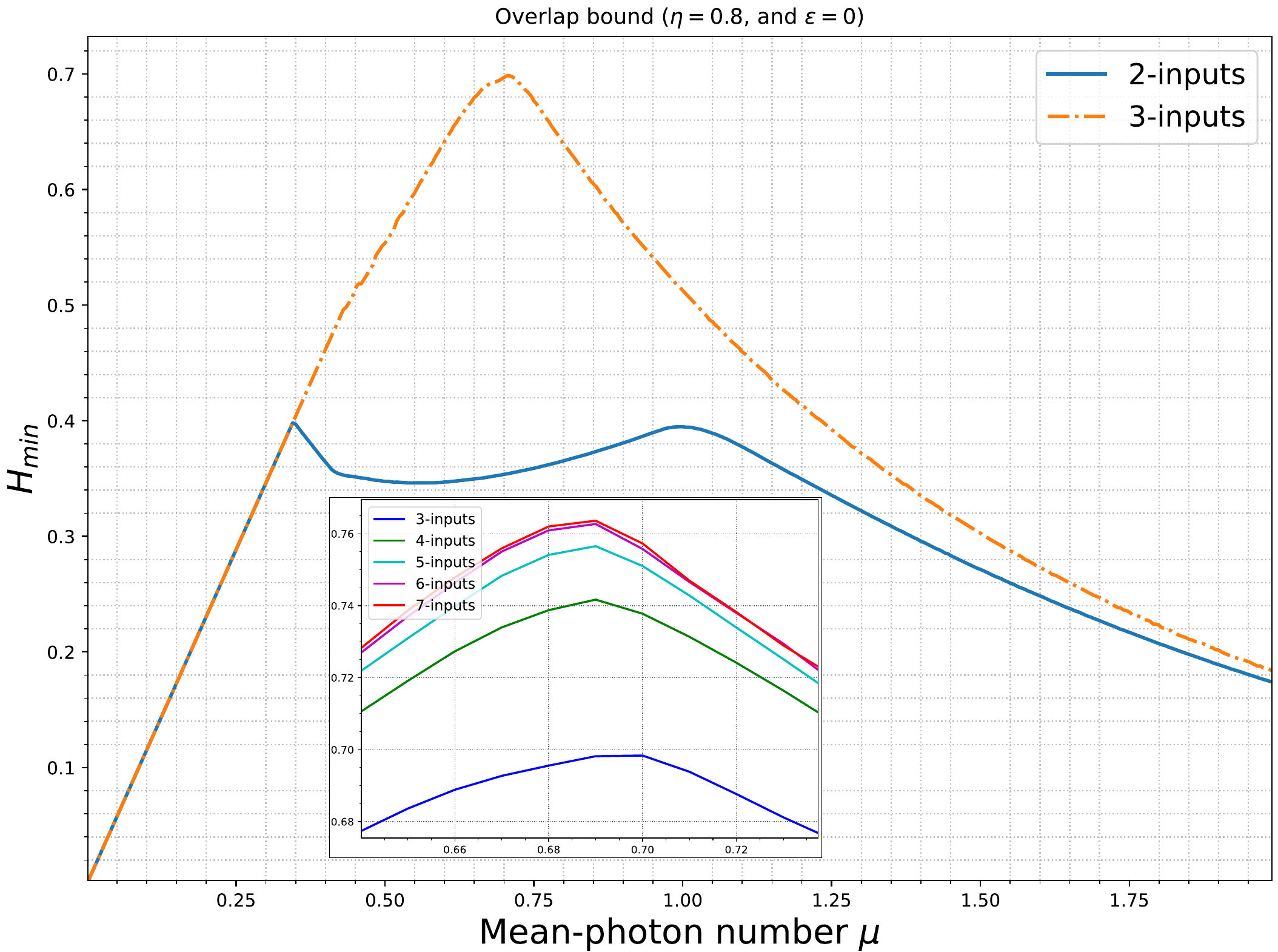}
\caption{The conditional min-entropy as a function of mean-photon number for a different number of inputs based on overlap assumption when the detector efficiency is $80\%$. 2-inputs describes two time-intervals; one is empty, the other has a weak-pulse (similar to \cite{Brask2017}), and 3-inputs is depicted in Fig. \ref{fig:general}. \textit{Inset}: more numbers of inputs is represented. }
\label{fig:compar}
\end{figure}
\subsection{Dual}
The dual SDP has three critical benefits when compared with the primal version:
it gives an upper-bound on the guessing probability rather than a lower-bound. In this way, conservative bounds are obtained, which never overestimates the min-entropy.  Further, the dual form enables recomputing bounds without running a full optimization for real-time operation, reducing the entropy estimation resources. Lastly, the finite-size effects can be
easily taken into account with this formulation.
Here, we use Lagrangian duality~\cite{boyd2004convex}, with an approach a similar to the one used in \cite{Bancal14,Brask2017}. 
We define the Lagrangian associated with the problem (\ref{eq:SDP_primal}) as:
\beq
\begin{aligned}
 {\mathcal L} = &\frac{1}{n} \sum\limits_{x = 0}^{n-1} \sum\limits_{\Lambda} \tr [{\rho _x}(\sum\limits_{b = 0}^{d-1}\delta_{\lambda_x,b} M _{b}^{\Lambda})] + \sum\limits_{\Lambda,b} {\tr[G_b^{\Lambda}M_b^{\Lambda}]}+ \\
& +\sum\limits_{\Lambda} {\tr[H^{\Lambda}\sum\limits_b {\left(M_b^{\Lambda} - \frac{1}{n} \tr[M_b^{\Lambda}]\right)} ]} +
\\
&+\sum\limits_{x,b} {{\nu _{bx}}\{\sum\limits_{\Lambda} {\tr[{\rho _x}M_b^{\Lambda}] - p(b|x)\}
} } \,,
\end{aligned}
\eeq
where $n\times n$ Hermitian matrices $ H^{\Lambda}$, $ G_b^{\Lambda}$, and scalar coefficient $\nu _{bx}$ are introduced as the Lagrange multipliers to each constraint in the primal problem. $\lambda_0,\dots, \lambda_{n-1}$ and $b$ range from $0$ to $d-1$, and $x$ ranges from $0$ to $n-1$. 
The next step is finding the supremum of the Lagrangian over the primal variables $M_b^{\Lambda}$. Now we minimize ${\mathcal L}$ over the Lagrangian multipliers to get a tighter bound on the guessing probability, so we have
\beq
\overbrace {\mathop {\sup ({\mathcal L} )}  \limits_{M_b^{\Lambda}} }^\chi = \mathop {\sup }\limits_{M_b^{\Lambda}} \{ \sum\limits_{\Lambda,b} {\tr[M_b^{\Lambda}J_b^{\Lambda}]} - \sum\limits_{x,b} {{\nu _{bx}}p(b|x)\} },
\label{SUP}
\eeq
where 
\beq
J_b^{\Lambda} = \sum\limits_x \rho _x(\frac{1}{n}\sum\limits_{b = 0}^{d-1}\delta_{\lambda_x,b}  + \nu _{bx}) 
+ G_b^{\Lambda} + H^{\Lambda}
- \frac{1}{n} \tr[H^{\Lambda}].
\label{JJ}
\eeq
Considering there is no constraint on $M_b^{\Lambda}$ in the Lagrangian, the supremum in Eq. (\ref{SUP}) will be infinite, except $J_b^{\Lambda}$ is restricted to be zero; thus we require that $J_b^{\Lambda}=0$. 

However, given that the operators $G_b^{\Lambda}$ are positive semidefinite, due to the second constraint of the primal SDP ($\ref{eq:SDP_primal}$), this is equivalent to cut $G_b^{\Lambda}$ from Eq. (\ref{SUP}) and expecting the rest of the expression to be negative semidefinite.
Consequently, we have the dualized SDP as
\beq
 P^*_g = \mathop {\min}\limits_{{H^{\Lambda}, \nu _{bx}}}  [- \sum_{x=0}^{n-1}\sum_{b=0}^{d-1} {{\nu _{bx}}p(b|x)}] 
\label{objec}
\eeq
subjected to

\begin{align}
&{H^{\Lambda}} = {\rm{ }}{({H^{\Lambda}})^\dag }, \\
& \sum\limits_x  {{\rho _x}(\frac{1}{n}\sum\limits_{b = 0}^{d-1}\delta_{\lambda_x,b} + {\nu _{bx}})} + {H^{\Lambda}} - \frac{1}{n} \tr[{H^{\Lambda}}]\mathbb{I} \le 0,
\label{const}
\end{align}
\section{Overlap bound and many inputs}
\label{app:c}
\begin{figure}[t!]
\centering
\includegraphics[width=\linewidth]{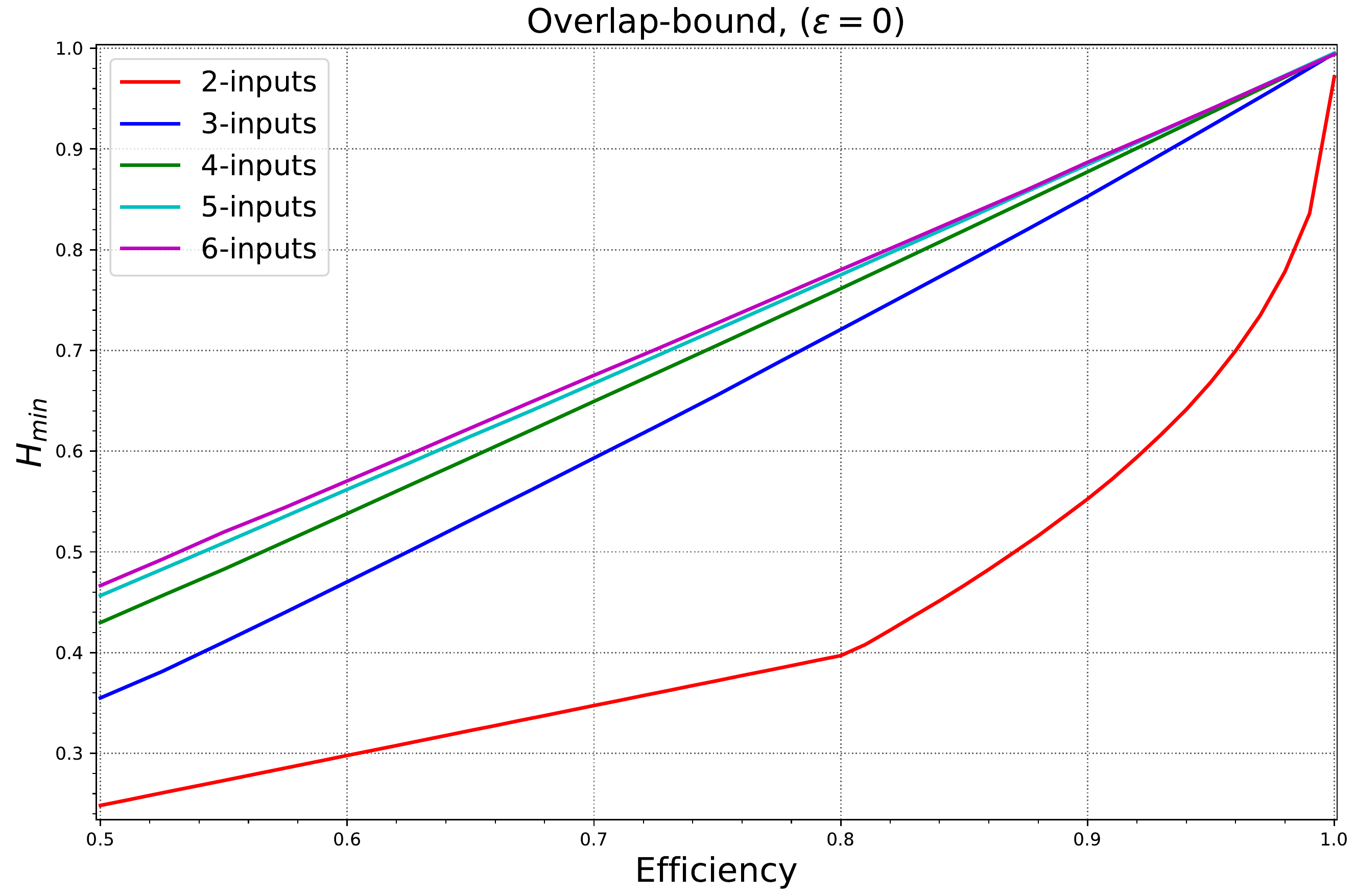}
\caption{The maximum achievable conditional min-entropy (with optimal mean-photon number) as a function of the detector's efficiency.}
\label{fig:efficiency}
\end{figure}
In this section we compare the energy bound considered so far 
$\langle \hat N\rangle_{\rho_x}\leq \mu$
with the
overlap bound assumption $\braket{\psi_x}{\psi_y}\geq\delta$ proposed in \cite{Brask2017}.
The advantage of the overlap bound assumption is that the QRNG could operate in a broader mean-photon number range and 
higher rates can be achieved.
However, from the experimental point of view, testing the energy bound is
easier than ensuring that the overlap bound is satisfied. 
We note that the bound on the energy imposes a bound on the overlap (see\cite{VanHimbeeck2017semidevice}), but not the other way around.
We will also compare the performances of the proposed implementation when the number of inputs are increased.

\subsection{Overlap bound}
To apply the overlap instead of the energy bound, we should change the assumption to 
\beq
|\braket{\psi_i}{\psi_j}|\geq e^{-\mu}, \;\;\;\;\;\; x,y \in \{0,1,2\}, \;\; x \neq y.
\label{Ovelap_bounddd}
\eeq
{For the estimation of the min-entropy with the overlap bound, we use the security framework described in the text (and in Appendix \ref{app:a}), with the only difference of the substitution of the overlap in Eq.\ref{state:ternary} with the one given by Eq. \ref{Ovelap_bounddd}. } 

In Fig. \ref{fig:compar}, the conditional min-entropy is plotted as a function of the mean-photon number for binary and ternary time-bin (Config. I) encoding schemes when the detector's efficiency is $80\%$. As shown, the maximum value of conditional min-entropy increases from $0.4$ to $0.7$, which is a significant improvement. 

We also show in the inset of Fig. \ref{fig:compar}
the numerical results obtained by
increasing the number of inputs to four, five, six and seven. It is worth to notice that, 
besides the extra experimental and computational complexity added by increasing the inputs, a negligible growth in the conditional min-entropy's maximum value is observed.  
Therefore, the ternary time-bin encoding scheme 
provides an excellent trade-off between the achievable conditional min-entropy and computational complexity.
It should be pointed out that when the number of inputs increases, the number of possible outcomes rise accordingly, and the guessing probability should be optimized over more measurement and preparation strategies. 
Thus, the optimization problem—either as a form of dual or primal SDP— would require more time to be determined, which reduces the system’s rate.
Notwithstanding, for a chosen number of input/output, the dual form can boost the generation rate compared to the primal form, 
since it allows to compute (sub-optimal) bounds without running a full optimization (the value of $P^*_g$ is linear in the experimental values $p(b|x)$.
We further show in Fig. \ref{fig:efficiency} the maximum achievable min-entropy (maximized of the possible $\mu$ values) in function of the detector's efficiency.
From the figure it is evident that increasing the number of outcomes from 2 to three increases the 
resistance to inefficiency.
As expected, the maximum achievable  min-entropy decreases by reducing the detector's efficiency, but only for 3 or more inputs it shows a quasi-linear behavior in function of the efficiency.

The gap between 2-inputs and 3-inputs cases grows when the detector efficiency decreases, while for the rest inputs, the gap is almost constant, see Fig. \ref{fig:efficiency}. This shows that the ternary encoding scheme is more robust to the detector efficiency than the binary one, which is an 
advantage as the typical single-photon detector's efficiency ranges from $0.5$ to $0.95$.
\begin{figure}[h!]
\centering
\includegraphics[width=\linewidth]{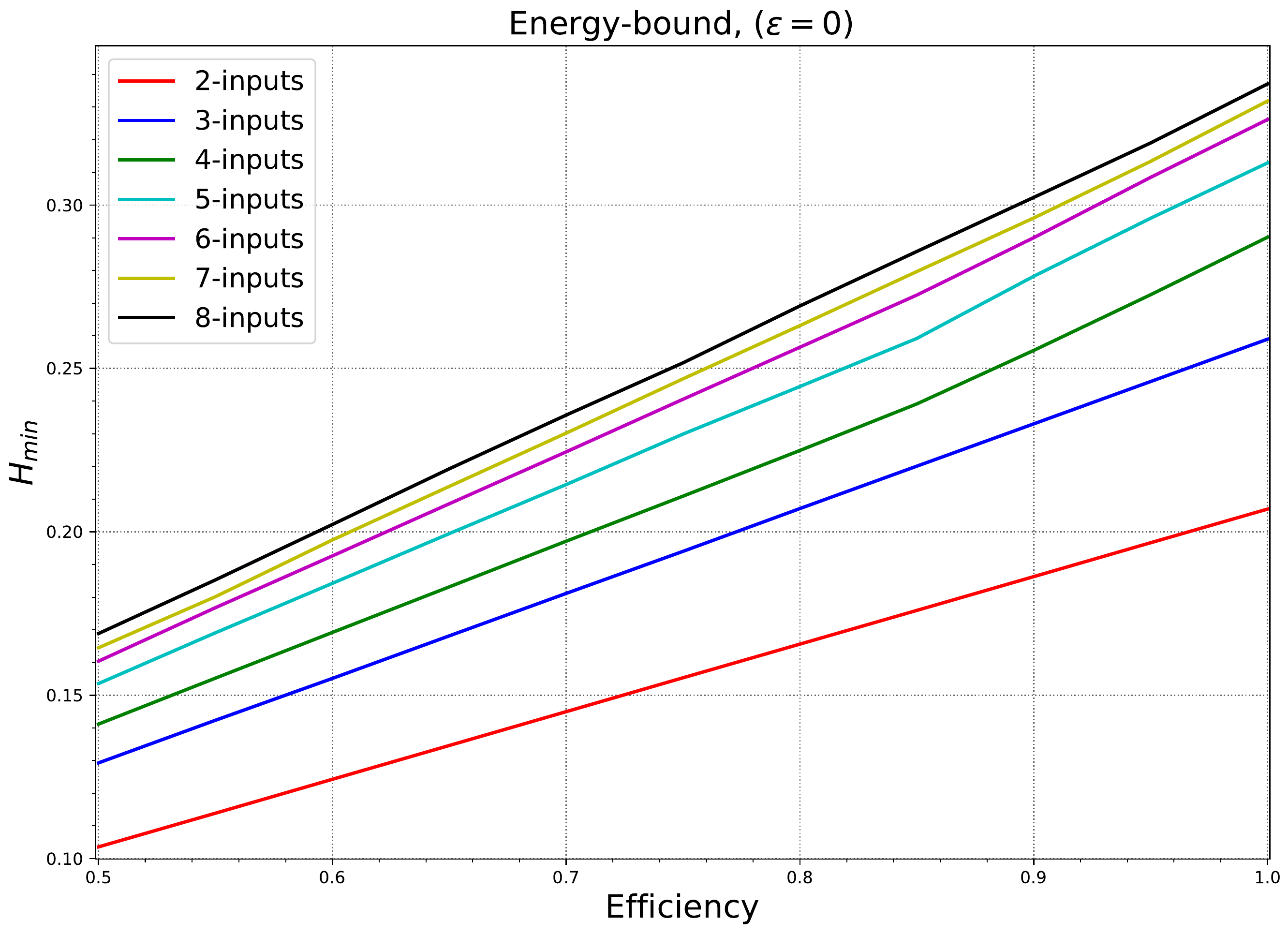}
\caption{The maximal achievable conditional min-entropy (with optimal mean-photon number) as a function of the detector's efficiency when the energy bound is considered.}
\label{fig:energy_assump}
\end{figure}
\subsection{Energy bound with many inputs}
\begin{figure}[h!]
\centering
\includegraphics[width=\linewidth]{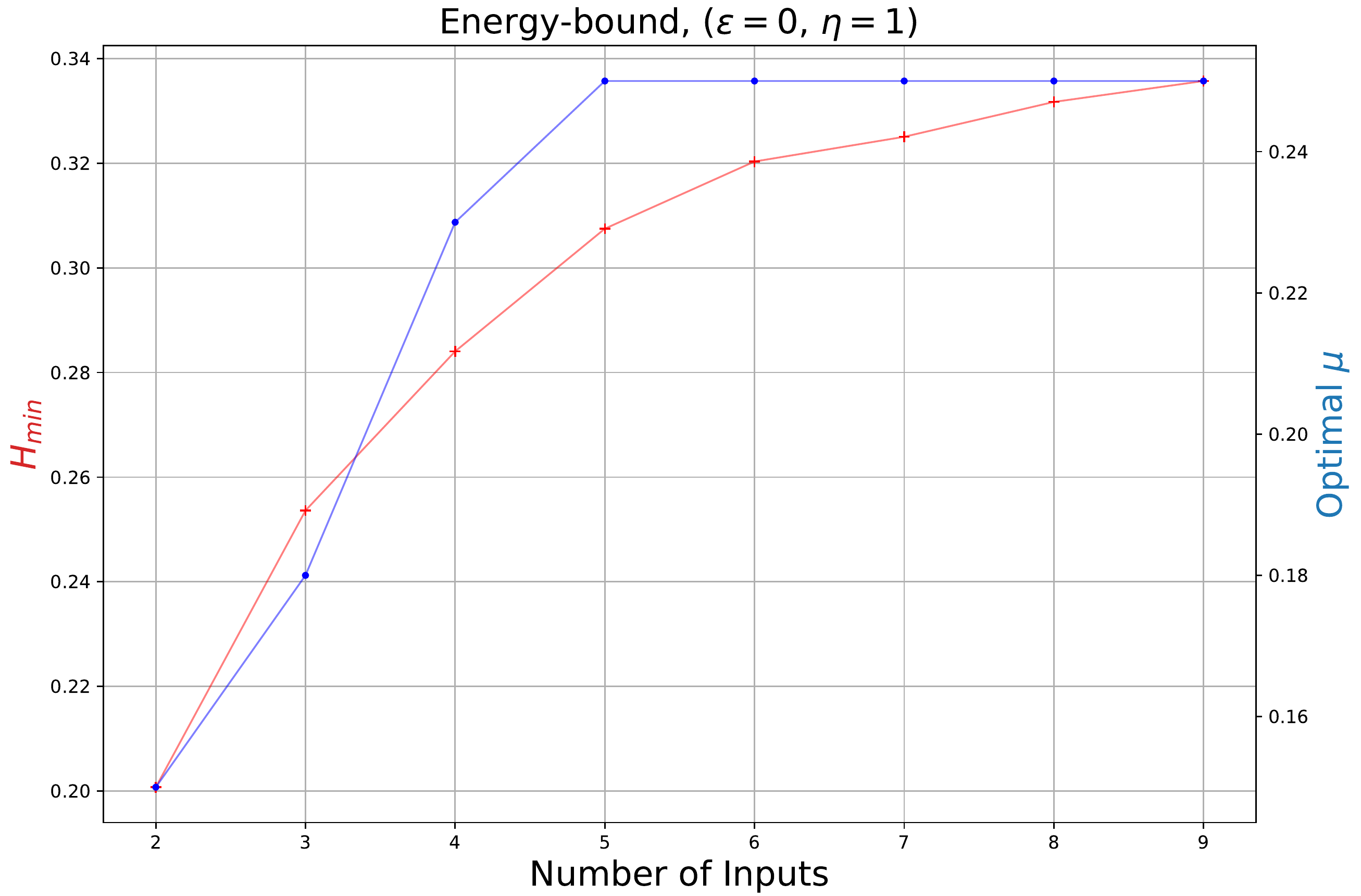}
\caption{The maximum conditional min-entropy $H_{min}$ and the corresponding optimal value of $\mu$  (for $\eta=1$) is plotted as a function of inputs. The value of the optimal $\mu$ raises when the number of inputs increases and asymptotically reaches a plateau of $ \sim 0.25$.}
\label{fig:optimal_mu}
\end{figure}
In this subsection, by employing the general SDP form given in the Appendix \ref{app:a}, 
we study the effect of changing the detector efficiency and the number of inputs when the energy bound is considered. 
Let's first consider the effect of detector efficiency, when no error are present ($\epsilon=0$). In Fig. \ref{fig:energy_assump} we show the maximum value of the min-entropy that can be achieved in function of the detection efficiency. 
Fig. \ref{fig:energy_assump} shows that increasing the number of inputs always improves the generation rate also when detection inefficiencies are taken into account. 
Consequently, it is possible to find the optimal trade-off between the computational complexity, entropy value, and robustness to the detector's efficiency.
Fig. \ref{fig:optimal_mu} shows the maximum min-entropy and the corresponding optimal value of $\mu$ as a function of the number of inputs in the
noiseless perfect-efficiency case ($\eta=1$, $\epsilon=0$). 
The data indicate that the optimal mean-photon number grows with the number of inputs and seemingly reaches a plateau of about $0.25$ for  high number of inputs ($>9$).
The 2-inputs results shown in Figs. (\ref{fig:efficiency}) and (\ref{fig:energy_assump}) illustrate that the binary inputs preparation scheme is less sensitive to the efficiency when  the energy bound is considered.
% \newpage
\end{document}